\documentclass[a4paper,11pt]{article}
\usepackage{blindtext}
\usepackage{geometry}
\usepackage[affil-it]{authblk}

\usepackage{graphicx} 
\usepackage{subfig}
\graphicspath{ {./images/} }
\usepackage{placeins}
\usepackage[rightcaption]{sidecap}
\usepackage[export]{adjustbox}
\usepackage{wrapfig}
\usepackage{amsmath}
\usepackage{tabularx}
\usepackage{hyperref}

\title{\textbf{Bubble dynamics in a cavitating venturi}}
\author[1,2*]{Premchand V Chandra}
\author[2,*]{Anuja Vijayan}
\author[2,*]{Pradeep Kumar}
\affil[1]{Indian Institute of Science Bangalore, India}
\affil[2]{Indian Institute of Space Science and Technology, Trivandrum, India}
 \affil[1,2,3]{(\textit{\textbf{*equal contributions}}; email: premchandv@iisc.ac.in)}
\begin{document}
\maketitle
\begin{abstract}
Cryogenic fluids are widely used as fuel for launch vehicles and many other industrial applications. The physics of cryogenic flows is highly complex due to their sensitivity to phase changes. Because of their low boiling points, cryogenic liquids undergo phase transition from liquid to bubbly liquid to vapour. This leads to cavitating flows that fall under the two-phase flow regime. This study models bubbly flows of cryogenic fluids, specifically liquid nitrogen, in a converging-diverging venturi device known as a cavitating venturi, a passive flow control and measurement device. Existing numerical studies in the literature are mostly limited to isothermal bubbly flows, such as water, where energy equations are often neglected due to minimal heat transfer at the interface.
In contrast, cryogenic fluids are highly sensitive to phase changes under ambient conditions and interface heat transfer plays a significant role owing to their low latent heat of vaporization. To develop an appropriate model for cryogenic bubbly flows, we incorporate the effects of convective heat transfer from the quiescent liquid to the traversing bubble and heat transfer at the bubble-fluid interface. The numerical modelling is carried out using an in-house finite-difference code, and the results highlight the critical influence of heat transport equations on bubble dynamics. A commercial CFD package is used for two-dimensional simulations to complement the in-house simulations to predict cavitation length, a key characterizing parameter. In addition, limited flow visualization experiments are performed using a high-speed camera to analyze the length of the cavitation zone. This combined numerical and experimental approach provides valuable insight into the dynamics of cryogenic bubbly flows in cavitating venturi systems.

\end{abstract}
\section{Introduction}
\begin{figure}
\centering
\includegraphics[scale=0.8]{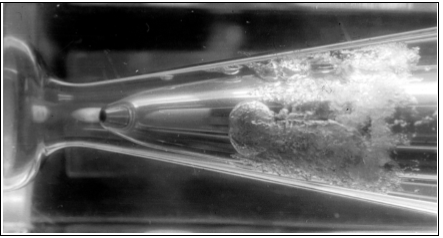}
\caption{Developed cavitation in a Venturi device \cite{1}}
\label{fig: Developed cavitation in a Venturi device}
\end{figure}

A cavitating venturi is a simple, passive flow control and metering device that maintains a constant mass flow rate of fluid, even when downstream pressure conditions fluctuate. The venturi consists of three key sections: a converging section, a throat, and a diverging section. The throat, having the minimum cross-sectional area, causes the fluid pressure to drop to its lowest point as the flow passes through it. When the fluid pressure near the throat reaches its vapour pressure, vapour cavities or bubbles begin to form, initiating cavitation \cite{1}. In a cavitating venturi, the flow exhibits a mix of behaviours, including bubble formation, growth, collapse, and dynamic interactions such as fission, fusion, re-entrant jet formation, and turbulent mixing of vapour bubbles as the flow moves from the throat into the diverging section. These characteristics are typical of cavitating two-phase flows. Lord Rayleigh's \cite{2} work was one of the earlier works on cavitation dynamics. Plesset et al. \cite{3} put forth a modified bubble dynamics equation for the present numerical work. Many theoretical, numerical, and experimental verification literature exists on cavitating flows and in particular the cavitating flows in converging diverging nozzles or cavitating venturi \cite{4}-\cite{10}.  Cavitating flow is a well-known two-phase phenomenon commonly encountered in mechanical systems such as turbo pumps (near the inducer) and marine propellers. In pumps operating at high rotational speeds, the pressure near the inducer can drop to levels that induce vapour formation, resulting in cavitation patterns such as attached cavitation and travelling bubble cavitation. These cavitation patterns are often observed in other turbomachines and can cause significant damage, including erosion of impeller vane surfaces, which ultimately shortens the service life.
Despite its detrimental effects on many mechanical systems, cavitation can benefit passive flow control devices such as cavitating venturi. The figure below (Figure \ref{fig: Developed cavitation in a Venturi device}) illustrates the development of cavitation in a typical cavitating venturi.
\\
\begin{figure}
\centering
\includegraphics[scale=0.6]{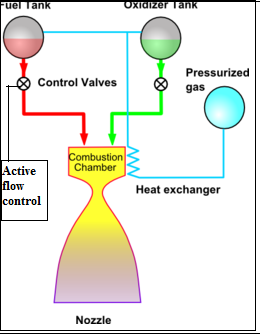}
 \includegraphics[scale=0.6]{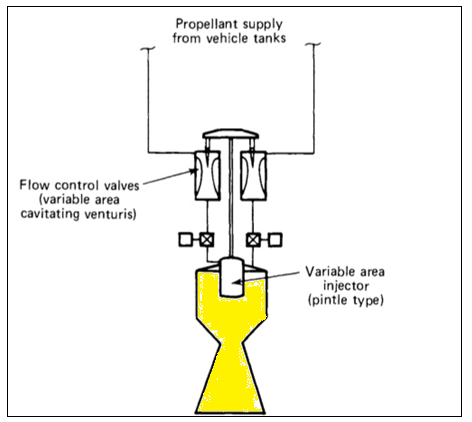}
\caption{Rocket Propulsion systems with (a). Active Flow device, and (b). Passive flow device}
\label{fig: Rocket Propulsion systems}
\end{figure}

As a passive flow measurement device, a cavitating venturi eliminates the need for complex control valves and actuating systems. Instead, it leverages the choked flow principle to maintain a constant flow rate, regardless of downstream pressure variations. Figure \ref{fig: Developed cavitation in a Venturi device} illustrates developed cavitation in a venturi.

Rocket propulsion systems rely on flow control to deliver liquid propellants to combustion chambers. While active systems involve control valves and actuators, their reliability is compromised by mechanical complexity. In contrast, a cavitating venturi serves as a reliable passive alternative, using cavitation-induced flow restriction to stabilize the flow rate.\\
In a cavitating venturi:\\
(a). For compressible flows, choked flow occurs as fluid accelerates to sonic velocity at the throat, preventing upstream pressure propagation.\\
(b). For incompressible flows, evaporation occurs before sonic velocity is reached, forming vapor bubbles near the throat. These bubbles restrict the flow, ensuring a stable flow rate.\\
(c). Downstream, pressure recovery collapses the vapor bubbles, restoring single-phase flow.\\

This unique property makes cavitating venturi ideal for liquid and cryogenic rocket engines, where flow remains steady despite fluctuations in combustion chamber pressure.
Cryogenic cavitating flows, due to their low latent heat of vaporization, exhibit complex behavior which mandates a detailed flow physics study. The key objectives of this work are:\\
1. To assess the predictive capabilities of one-dimensional models for cryogenic cavitating venturis. Though simplified, 1D models provide valuable estimates of key flow parameters.\\

2. To develop two-dimensional models using ANSYS Fluent with appropriate cavitation models. Also to predict flow characteristics, such as cavitation length, and validate against experiments.\\

3. Design and develop a planar venturi test facility with flow visualization features. Conduct experiments using cryogenic liquid nitrogen to measure cavitation length and flow behavior.\\
This study builds upon Wang and Brennen's approach \cite{9}, extending it to include thermal effects in cryogenic flows. Numerical simulations with 2D models provide a robust framework to capture the complex dynamics of cavitating venturi systems.

\section{Numerical 1D model}
In Brennen \cite{9} model, the Rayleigh Plesset Equation for the isothermal case was assumed for solving the bubble dynamics. However, this bubble dynamics equation cannot be extended to the case of cryogenic liquids, as heat transfer effects are predominant in cryogenic liquids. It is advantageous to have an additional term in the same Brennen model \cite{9} that accounts for heat transfer and extends to cryogenic cavitating venturi flows. The following section details models used for the Numerical 1D studies for cryogenic cavitating flows.

\subsection{Derivation of modified Rayleigh-Plesset equation}

The following derivation systematically introduces the thermal term into the Bubble dynamics equation. Starting from the classical form of the Rayleigh Plesset equation, the derivation for the modified Rayleigh Plesset equation is as follows

\begin{equation}\label{eq:a}
\left[ R \ddot{R} + \dfrac{3}{2} \dot{R}^{2} \right] = \dfrac{\left[ P_{\nu}-P_{\infty}(t) \right] }{\rho_{L}} + \dfrac{P_{g0}}{\rho_{L}}\left[ \dfrac{R_{0}}{R} \right]^{3 \gamma} - \dfrac{2S}{\rho_{L}R} - \dfrac{4\mu}{\rho_{L}}\dfrac{\dot{R}}{R}
\end{equation}

The above is the simple Rayleigh-Plesset equation for the isothermal case. The equation does not have an appropriate term for handling the heat interactions. The vapour pressure inside the bubble can be represented as $P_{v}(T_{B})$ in this equation. In an iso-thermal fluid such as water, there is no temperature difference between the bubble temperature $T_{B}$  and the surrounding liquid temperature $T_{\infty}$. The above isothermal condition is an assumption which is not valid for cryogenic liquids as the slightest temperature difference between $T_{B}$  and $T_{\infty}$ causes a considerable change in the vapour pressure inside the bubble. The temperature gradient $T_{\infty} - T_{B}$  becomes a potential source for heat transfer influencing the bubble growth.

 By small algebraic manipulation of adding and subtracting $ P_{\nu}(T_{\infty}) $ in $ \left[ P_{\nu}(T_{B})-P_{\infty}(t) \right] $  it can be written as

\begin{equation} \label{eq:21}
\left[ P_{\nu}(T_{B})-P_{\infty}(t) \right] = \left[ P_{\nu}(T_{\infty})-P_{\infty}(t) \right]+ \left[ P_{\nu}(T_{B})-P_{\nu}(T_{\infty}) \right]
\end{equation}
Substituting equation \ref{eq:21} in \ref{eq:a} the equation becomes of this form
\begin{equation} \label{eq:22}
\left[ R \ddot{R} + \dfrac{3}{2} \dot{R}^{2} \right] +\dfrac{2S}{\rho_{L}R} +\dfrac{4\mu}{\rho_{L}}\dfrac{\dot{R}}{R} =  \dfrac{P_{g0}}{\rho_{L}}\left[ \dfrac{R_{0}}{R} \right]^{3\gamma} + \dfrac{\left[ P_{\nu}(T_{\infty})-P_{\infty}(t) \right]}{\rho_{L}} - \dfrac{\left[ P_{\nu}(T_{B})-P_{\nu}(T_{\infty}) \right]}{\rho_{L}}
\end{equation}

 The third term on the right-hand side of the equation is the thermal term, and the second term is the driving term for the bubble growth. To find an expression for the thermal term, we need to look into the heat Balance and B-factor theory \cite{1} \& \cite{5}
\\
 The thermal effect in cryogenic liquids can be interpreted as the heat interaction between the vapour bubbles and the surrounding liquid. The production of a vapour volume $V_{\nu}$   requires the quantity of heat $\rho_{\nu}V_{\nu}L $  to be supplied (L is the latent heat of vaporization). This energy is given by the liquid near the vapour cavity. Hence, the liquid mass surrounding the vapour bubble gets cooled by supplying heat to the bubble. Steffano \cite{1} \& \cite{5} put forth a theory, popularly known as B-factor theory, which gives a non-dimensional number for temperature drop called B-factor \cite{1}
\\
 The volume of liquid $V_{L}$ after supplying the heat of vaporization reduces itself with its temperature by a drop $\Delta T$, and the heat balance between the liquid and vapour is given by
\begin{equation} \label{eq:23}
\rho_{\nu}V_{\nu}L=\rho_{l}V_{l}C_{pl}\Delta T
\end{equation}

 From the above equation, the B-factor gives the ratio of the volume of vapour to that of the volume of liquid. Alternately, it is also given by the ratio of the actual temperature drop to the reference temperature drop, which is shown below 
\begin{equation} 
B=\dfrac{\Delta T}{\dfrac{\rho_{\nu}L}{\rho_{l}C_{pl}}}=\dfrac{V_{\nu}}{V_{l}}
\end{equation}
Where,\\
 $ \Delta T^{\ast} = \dfrac{\rho_{\nu}L}{\rho_{l}C_{pl}} $ is the reference temperature drop and $ \Delta T $ is the actual temperature drop. 
\\ \\
This slight temperature drop $ \Delta T $ in the liquid happens to be the reason for the difference in temperature between a vapour bubble and the liquid in the immediate vicinity, which affects the vapor pressure change inside the bubble, as the bubble temperature has now been slightly changed. If the drop in temperature is very small, then by Taylor series expansion, we can write
\begin{equation} 
\Delta P_{\nu}=\dfrac{dP_{\nu}}{dT}\Delta T
\end{equation}

The vapour pressure changes with respect to the temperature $ \dfrac{dP_{\nu}}{dT} $ which is given by the Clausius Clayperon relation\cite{8}

\begin{equation} 
L=T\left[ \dfrac{1}{\rho_{\nu}} -\dfrac{1}{\rho_{l}} \right] \dfrac{dP_{\nu}}{dT}
\end{equation}

Usually, the liquid density $ \rho_{l} $ in the Clausius-Clapeyron relation is neglected, and the relation reduces to
\begin{equation} 
L=\dfrac{T_{\infty}}{\rho_{\nu}}\dfrac{dP_{\nu}}{dT}
\end{equation}

To estimate the temperature difference $ \Delta T $, the heat transfer between liquid and vapour bubble has to be quantified. In literature, both conductive and convective modes of heat transfer were assumed based on the flow conditions, i.e., if there is no slip between liquid and bubble, then the conductive approach is sufficient. But in our case, assuming a convective mode of heat transfer is applicable.
\\ \\
 Assuming $Q$ as the heat flux at the interface of bubble and liquid, then we can write
\begin{equation} 
Q=hA\delta T
\end{equation}
As this heat supplied by the liquid is the actual heat source for vapourization and growth of the bubble, a balance of heat can be done as shown below 
\begin{equation} 
h\delta T (4\pi R^{2})=\dfrac{d}{dt}\left( \dfrac{4}{3}\pi R^{3} \right) \rho_{\nu} L
\end{equation}

 Rearranging the equation in terms of $ \delta T $ gives,
\begin{equation} 
\delta T = \dfrac{\rho_{\nu} L}{h}\dot{R}
\end{equation}

Positive $ \delta T $  indicates a temperature decrement in liquid, but a temperature increment in vapour bubble. This, in turn, increases the vapour pressure inside the bubble, which in turn makes the bubble expand or grow. Similarly, it can be interpreted that the negative $ \delta T $  leads to shrinkage of the vapour bubble due to condensation.
\\ \\
 Now, coming back to the thermal term as stated in equation \ref{eq:22}, we can write 
\begin{center}
$ \dfrac{\left[ P_{\nu}(T_{B})-P_{\nu}(T_{\infty}) \right]}{\rho_{L}} = \dfrac{\delta P_{\nu} }{\rho_{L}}$
\end{center}

substituting  $ \delta P_{\nu}  $  from equation[25] gives
\begin{center}
$ \dfrac{\left[ P_{\nu}(T_{B})-P_{\nu}(T_{\infty}) \right]}{\rho_{L}} = \dfrac{dP_{\nu} }{dT} \dfrac{\delta T}{\rho_{L}}$
\end{center}

 again $ \delta T $ can be substituted from equation[29] which results in an expression of thermal term as follows
\begin{equation} 
\dfrac{\left[ P_{\nu}(T_{B})-P_{\nu}(T_{\infty}) \right]}{\rho_{L}} =  \dfrac{dP_{\nu} }{dT} \dfrac{\rho_{\nu}L}{\rho_{L}}\dfrac{\dot{R}}{h}
\end{equation}

After substituting equation[30] into [22] we finally get the \textbf{modified Rayleigh-Plesset} equation with thermal source term as follows
\begin{eqnarray}\label{eq:31}
\left[ R \ddot{R} + \dfrac{3}{2} \dot{R}^{2} \right] +\dfrac{2S}{\rho_{L}R} +\dfrac{4\mu}{\rho_{L}}\dfrac{\dot{R}}{R} =  \dfrac{P_{g0}}{\rho_{L}}\left[ \dfrac{R_{0}}{R} \right]^{3\gamma} + \dfrac{\left[ P_{\nu}(T_{\infty})-P_{\infty}(t) \right]}{\rho}_{L} - \dfrac{dP_{\nu}}{dT}\dfrac{\rho_{\nu} L}{\rho_{L}}\dfrac{\dot{R}}{h}
\end{eqnarray}

Equation \ref{eq:31} is the governing differential equation for bubble dynamics applicable for non-isothermal liquids such as cryogenic liquids.

\subsection{Modelling 1D cavitating nozzle / Venturi flow}
\paragraph{}
\indent  Similar to Wang and Brennen Quasi-1D model \cite{9} for Bubbly Cavitating flows through a nozzle, a Quasi-1D model with thermal effects is attempted in this work applicable to cryogenic liquids. The continuity and momentum equations for bubbly two-phase flows along with the closures, i.e., the volume fraction equation and the modified Rayleigh Plesset equation along with the bubble momentum equation for estimation of bubble velocity, were considered in this model for solving bubble radius.
\\ \\
\indent For modelling the heat transfer between bubble and liquid, a convective heat transfer model was taken from the existing Ranz and Marshall model \cite{14}. The turbulence model was not incorporated in 1D numerical studies, as handling such equations would be tedious. Following assumptions were made in this model,
\begin{enumerate}
\item At any cross-section of the venturi, the bubbles had a uniform size, and the bubble number was assumed constant.
\item The bubble and the liquid temperatures were equal at the interface, and there was no friction or heat transfer between the venturi walls and the flowing fluid.
\item Liquid and vapour densities were assumed to be constant
\end{enumerate}

\indent Supposing that flow develops only in the longitudinal x direction, the steady bubbly flow continuity and momentum equations that Wang and Brennen used are given as follows
\begin{center}
$\dfrac{\partial (1-\alpha)A}{\partial t} + \dfrac{\partial (1-\alpha)Au}{\partial x} = 0$\\
$ \dfrac{\partial u}{\partial t} + u \dfrac{\partial u}{\partial x} =- \dfrac{1}{2(1-\alpha)} \dfrac{C_{p}}{\partial x} $
\end{center}

\indent Where, $ C_{p}(x,t)=\dfrac{(P(x,t)-P_{0})}{(0.5\rho_{l}u_{0}^{2})} $  is the fluid pressure coefficient $ P(x,t) $  is the fluid pressure , $ P_{0} $ is the upstream fluid pressure, $ u_{0} $  is the upstream fluid velocity, and $ u $ is the velocity along the length of the profile. $A(x)$ corresponds to the area profile, which is the function of $x$ (being a 1D case). The mathematical expression for the area profile A(x) is
\begin{equation}{A(x) =}
	 \left\lbrace  1 - \dfrac{1}{2} C_{p,min} \left[ 1 - cos \left( \dfrac{2 \pi x}{L} \right)  \right] \right\rbrace^{- \dfrac{1}{2}} 
\end{equation}

\indent $ \alpha(x,t)=\dfrac{\dfrac{4}{3}\pi \eta R(x,t)^{3}}{\left( 1+\dfrac{4}{3} \pi \eta R(x,t)^{3} \right) } $ is the non-dimensional void fraction, $R(x,t)$ is the non-dimensional bubble radius, and $ \eta $ is the non-dimensional population number.
\\ \\

\indent For modelling the non-linear bubble dynamics, a modified Rayleigh–Plesset equation \ref{eq:31} was used
\begin{equation}
\left[ R \ddot{R} + \dfrac{3}{2} \dot{R}^{2} \right] +\dfrac{2S}{\rho_{L}R} +\dfrac{4\mu}{\rho_{L}}\dfrac{\dot{R}}{R} =  \dfrac{P_{g0}}{\rho_{L}}\left[ \dfrac{R_{0}}{R} \right]^{3\gamma} + \dfrac{\left[ P_{\nu}(T_{\infty})-P_{\infty}(t) \right]}{\rho_{L}} - \dfrac{dP_{\nu}}{dT}\dfrac{\rho_{\nu} L}{\rho_{L}}\dfrac{\dot{R}}{h}
\end{equation}

\indent The above equation can be represented in total derivative form by replacing $ \dot{R}=\dfrac{DR}{Dt} $, which is given by
\begin{multline}
\left[ R \dfrac{D^{2}R}{Dt^{2}} + \dfrac{3}{2} \left( \dfrac{DR}{Dt} \right)^{2}  \right] +\dfrac{2S}{\rho_{l}R} +\dfrac{4\nu_{l}DR}{RDt} =  \dfrac{P_{g0}}{\rho_{l}}\left[ \dfrac{R_{0}}{R} \right]^{3\gamma} 
\\
+ \dfrac{\left[ P_{\nu}(T_{\infty})-P(x,t) \right]}{\rho_{L}} - \dfrac{dp_{\nu}}{dT}\dfrac{\rho_{\nu} L_{e\nu}}{\rho_{L}h_{b}}\dfrac{DR}{Dt}
\end{multline}

\indent In the above equation, the last term on the right side is the thermal term, and  $ h_{b} $ is the convective heat transfer coefficient, which was modelled using the already existing Ranz and Marshall \cite{14} model for the study of evaporation of droplets in spray drying which is given by
\begin{center}
$ h_{b} = \dfrac{Nu_{b} K_{l}}{2R} $
\end{center}

\indent Where  $ Nu_{b} = 2 + 0.6 Reb^{\frac{1}{2}}Pr^{\frac{1}{3}} $ is the bubble Nusselt number, which in turn depends on the bubble Reynolds number ($ Re_{b} $), and Prandtl number (Pr) which is given by \cite{8} as
\begin{center}
$ Re_{b} = 2R\vert \nu- u \vert /\lambda_{l} $\\

$ Pr = \dfrac{C_{pl}\mu_{l}}{K_{l}} $
\end{center}

\indent where  $ \lambda_{l} = K_{l}/\rho C_{p} $ is the thermal diffusivity and $ K_{l} $  is the liquid thermal conductivity.
\\ \\
\indent For solving bubble velocity, the bubble momentum equation formulated by Albagli et al. \cite{13} was used, which is shown as follows
\begin{center}
$ \rho_{\nu}\dfrac{D\nu}{Dt} + \dfrac{1}{2}\rho_{l}\left[ \dfrac{D\nu}{Dt}-\dfrac{Du}{Dt} \right] = -\dfrac{\partial P(x,t)}{\partial x} - \dfrac{3}{8}\rho_{l}C_{D} \times \dfrac{(\nu-u)\vert \nu-u\vert}{R} $
\end{center}

\subsection{Wang and Brennen benchmark Results for Water with thermal term}
\paragraph{}

\begin{figure}[ht!]
\centering
\includegraphics[scale=0.45]{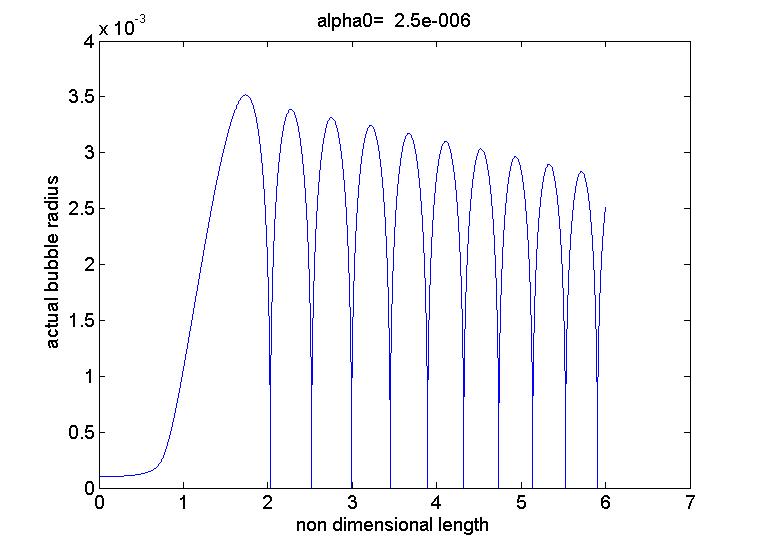}
\caption{Non-dimensional Bubble radius plot for water at  $\alpha_{0} = 2.5 \times 10^{-6}$ }
\label{fig:chp3fig2}
\end{figure}

\begin{figure}[ht!]
\centering
\includegraphics[scale=0.45]{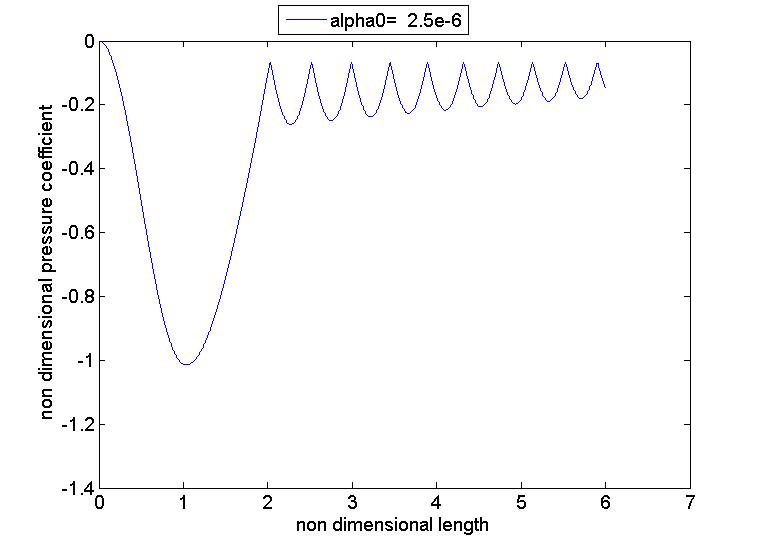}
\caption{Coefficient of Pressure  plot for water at   $\alpha_{0} = 2.5 \times 10^{-6}$ }
\label{fig:chp3fig6}
\end{figure}

\begin{figure}[ht!]
\centering
\includegraphics[scale=0.45]{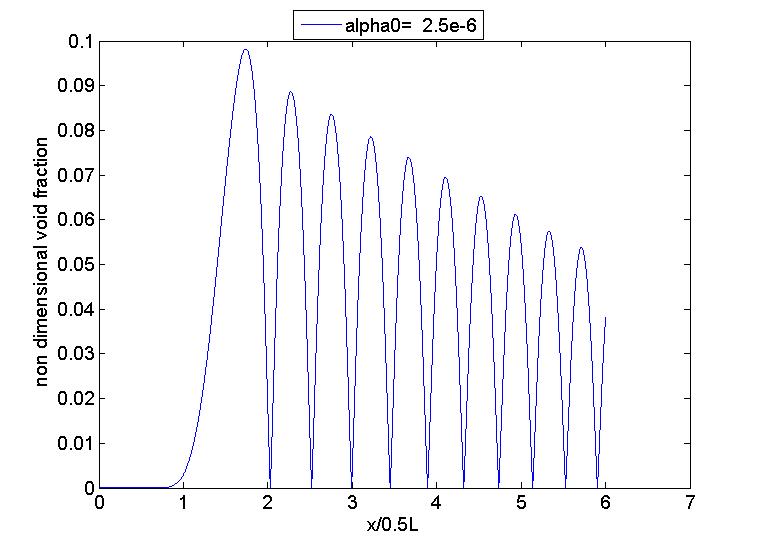}
\caption{Void fraction plot for water at    $\alpha_{0} = 2.5 \times 10^{-6}$ }
\label{fig:chp3fig7}
\end{figure}

\begin{figure}[ht!]
\centering
\includegraphics[scale=0.45]{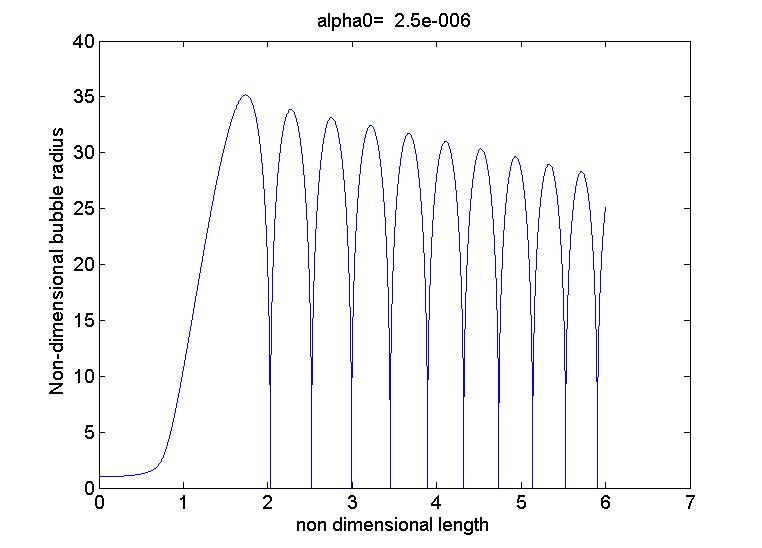}
\caption{Non-dimensional bubble radius plot for water at   $\alpha_{0} = 2.5 \times 10^{-6}$ }
\label{fig:chp3fig8}
\end{figure}

\indent A numerical 1D code using Matlab was written for solving the Brennen model with water as the fluid considering the effects of the thermal term.  The results were plotted for the bubble radius, void fraction, coefficient of pressure, velocity of the liquid phase, and the growth contribution terms (thermal and pressure terms). The non-dimensional bubble radius, coefficient of pressure, void fraction, and non-dimensional void fraction plots are shown below in the figure \ref{fig:chp3fig2},\ref{fig:chp3fig6}, \ref{fig:chp3fig7}, \ref{fig:chp3fig8}corresponds to the Rayleigh Plesset bubble dynamics equation with the thermal effects incorporated.

\begin{figure}[ht!]
\centering
\includegraphics[scale=0.55]{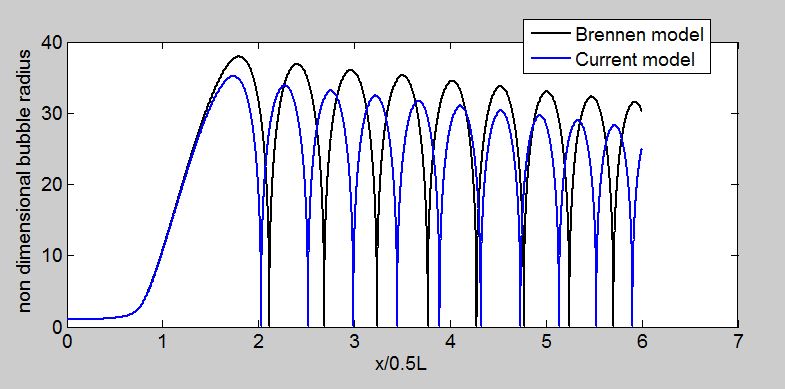}
\caption{Non-Dimensional Bubble radius plot from Wang $\&$ Brennen \cite{9} model and current model }
\label{fig:chp3fig3}
\end{figure}

\begin{figure}[ht!]
\centering
\includegraphics[scale=0.45]{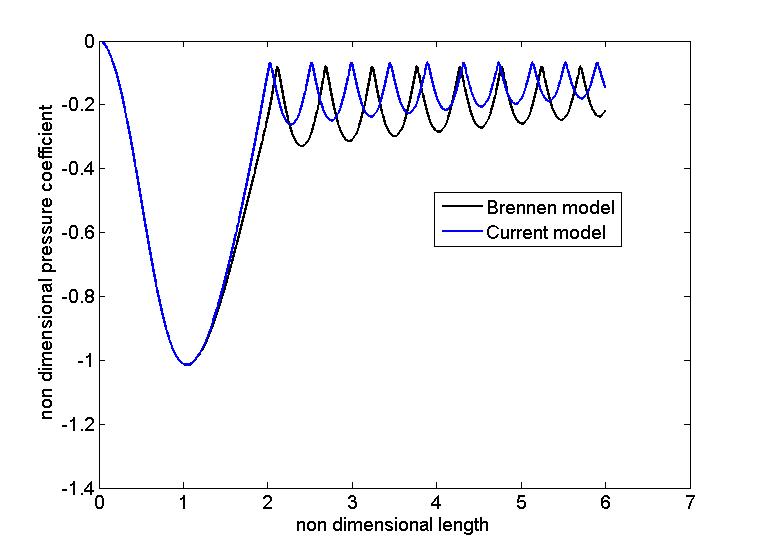}
\caption{Coefficient of Pressure plot from Wang $\&$ Brennen \cite{9} model and current model }
\label{fig:chp3fig4}
\end{figure}

\begin{figure}[ht!]
\centering
\includegraphics[scale=0.45]{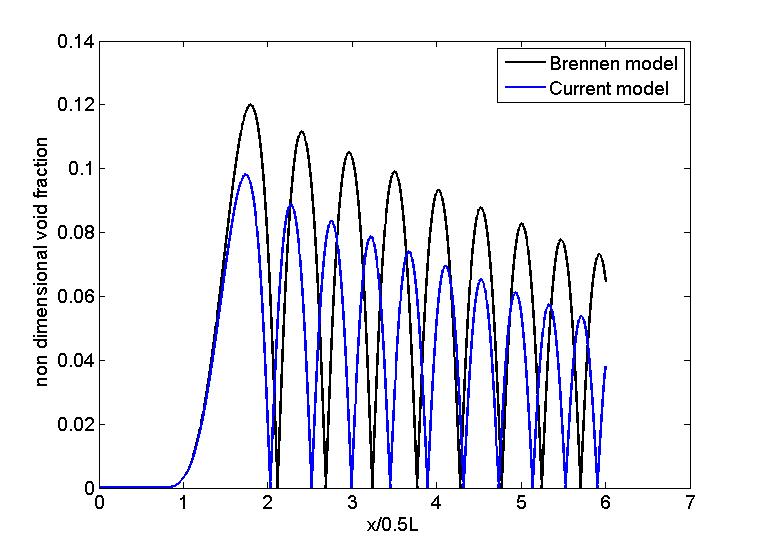}
\caption{Void fraction plot from Wang $\&$ Brennen \cite{9} model and current model }
\label{fig:chp3fig5}
\end{figure}

\indent The superimposed plot of Wang and Brennen \cite{9} bubble radius, coefficient of pressure, and void fraction plot with the current numerical model(with thermal effects) plot is shown in the figure \ref{fig:chp3fig3},\ref{fig:chp3fig4},\ref{fig:chp3fig5}

\indent It can be inferred from the superimposed figures \ref{fig:chp3fig3},\ref{fig:chp3fig4}, \ref{fig:chp3fig5} that there is no much variation in the bubble radius, void fraction, and coefficient fraction plot of current model and Brennen model \cite{9}. However, deviations may be attributed to the additional thermal effect term. This indicated that including the thermal effect in the Rayleigh Plesset equation has no considerable role to play in the bubble dynamics of Water. The reason for this is the minimal contribution of Thermal term in water.
\\ 

\begin{figure}[ht!]
\centering
\includegraphics[scale=0.5]{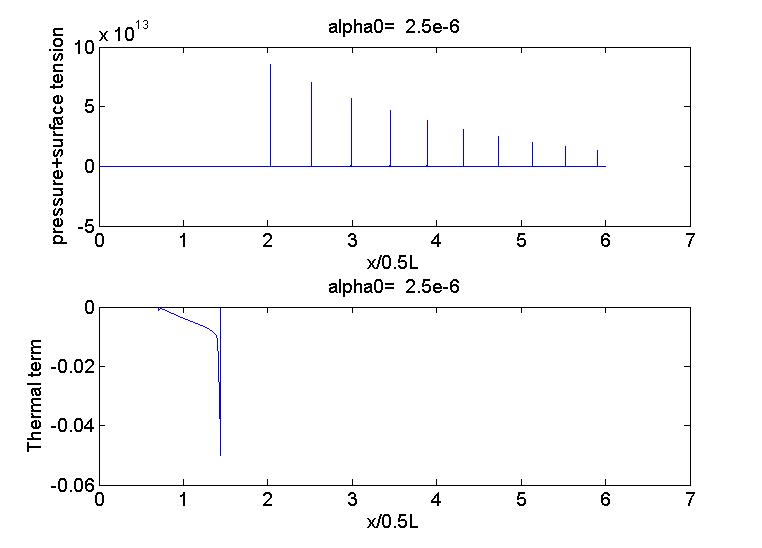}
\caption{Source term plots for water at    $\alpha_{0} = 2.5 \times 10^{-6}$ }
\label{fig:chp3fig9}
\end{figure}

\indent The figure \ref{fig:chp3fig9} gives the plot of source terms in the Rayleigh-Plesset equation. The thermal source term is compared with a pressure source term. The values of the same are plotted for an initial void fraction of $\alpha_{0} = 2.5 \times 10^{-6}$, where the term $\alpha_{0}$ is the initial void fraction which has been arbitrarily assumed in the Wang and Brennen model \cite{9}. This initial void fraction significantly impacts the numerical prediction of bubble radius as this signifies the number of nucleation sites or microscopic voids in the liquid phase.
\\ 
\indent For understanding the domination of different source terms in the bubble dynamics, the right-hand side of the modified Rayleigh-Plesset equation containing such source terms was handled separately, and the magnitude of the same were plotted. i.e., all other terms except the thermal term were taken as the pressure source term. The following figure \ref{fig:chp3fig9},\ref{fig:chp3fig10} shows the contribution of the above-mentioned thermal term and pressure term.

\begin{figure}[ht!]
\centering
\includegraphics[scale=0.5]{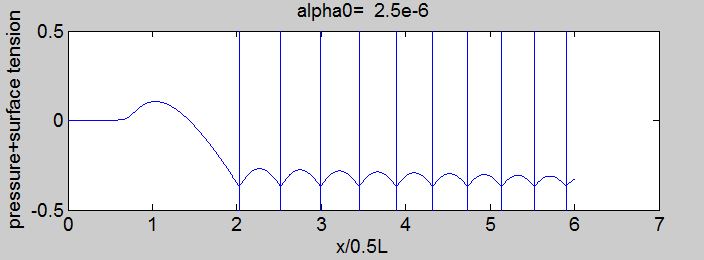}
\caption{Magnified view of pressure source term plots for water at    $\alpha_{0} = 2.5 \times 10^{-6}$ }
\label{fig:chp3fig10}
\end{figure}

\indent It can be inferred from the plots that the contribution of the thermal term within the flow domain of the venturi is small and is of the order $10^{-1}$ when compared to the contribution from the pressure term, which is of the order $10^{4}$. This points out to the fact that the explicit modelling of the thermal term in the modified Rayleigh Plesset Equation may not provide pronounced effects for fluids with the high latent heat of vapourization. This fact explains the close prediction of the Wang $\&$ Brennen model results , where such a thermal term is not modelled. 

 Having benchmarked the model, the same model is now extended to predict cavitating flows of liquid nitrogen. 

 However, the present one-dimensional model is far from providing estimates for performance prediction of cavitating venturi, and more systematic work in developing proper mathematical models to predict the performance would be required. Hence, a systematic study of a simple planar 2D numerical study was attempted using ANSYS fluent, which will be discussed.
\subsection{Numerical 1D Results: for the present case}
\paragraph{}

The numerical 1D simulation was done for liquid nitrogen, including the thermal effect, using the modified Rayleigh Plesset equation using MATLAB.

\indent Bubble radius, coefficient of pressure, and void fraction were plotted for different cases. Since the numerical model is an initial value problem, the initial values were assumed for all the variables in line with Wang and Brennen \cite{9} model. It was inferred that the initial void fraction had a considerable effect on the nature of the solution.

\subsection{1D Numerical Test cases}
\subsubsection{Case-1}
\paragraph{}
 For an Initial void fraction $ \alpha_{0} = 2.5 \times 10^{-6} $ and $C_{p}$=0\\
 
\begin{figure}[ht!]
\centering
\includegraphics[scale=0.45]{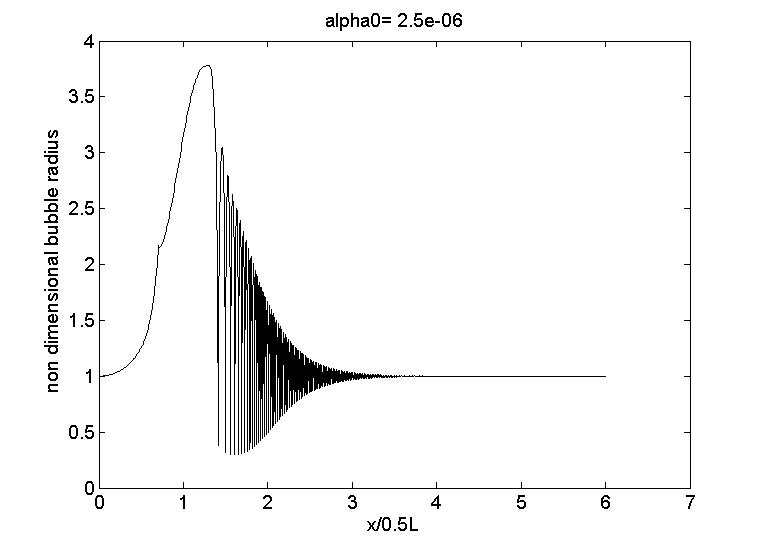}
\caption{Non-dimensional bubble radius plot for case 1 of Liquid Nitrogen}
\label{fig:chap5fig1}
\end{figure}

\begin{figure}[ht!]
\centering
\includegraphics[scale=0.45]{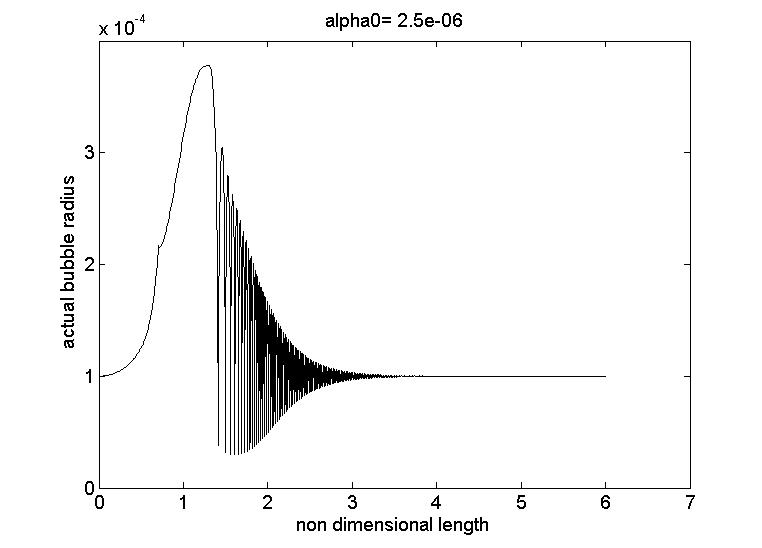}
\caption{Actual bubble radius plot for case 1 of Liquid Nitrogen}
\label{fig:chap5fig2}
\end{figure}

 From the figure \ref{fig:chap5fig1},\ref{fig:chap5fig2}, it is evident that the non-dimensional bubble radius grows along the converging section from $1$ to $3.4$ up to the throat($x=0$ indicates the beginning of the converging section as per Brennen Area profile,$x=1$ indicates the throat section (or) beginning of the diverging section and $x=2$ indicates the exit of diverging section), and dampens from the throat to the initial value of 1 at the exit of the nozzle and the region away from the nozzle exit.

\begin{figure}[ht!]
\centering
\includegraphics[scale=0.45]{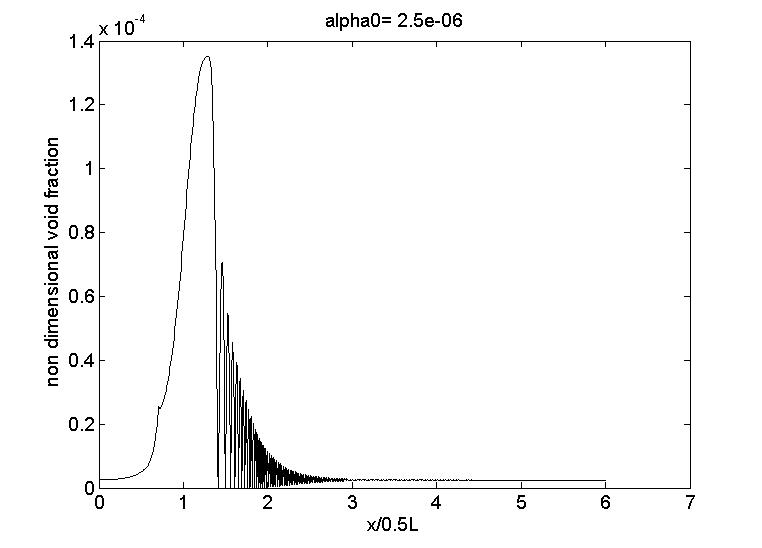}
\caption{Non dimensional Void fraction plot for the case1 of Liquid nitrogen}
\label{fig:chap5fig3}
\end{figure}

\begin{figure}[ht!]
\centering
\includegraphics[scale=0.4]{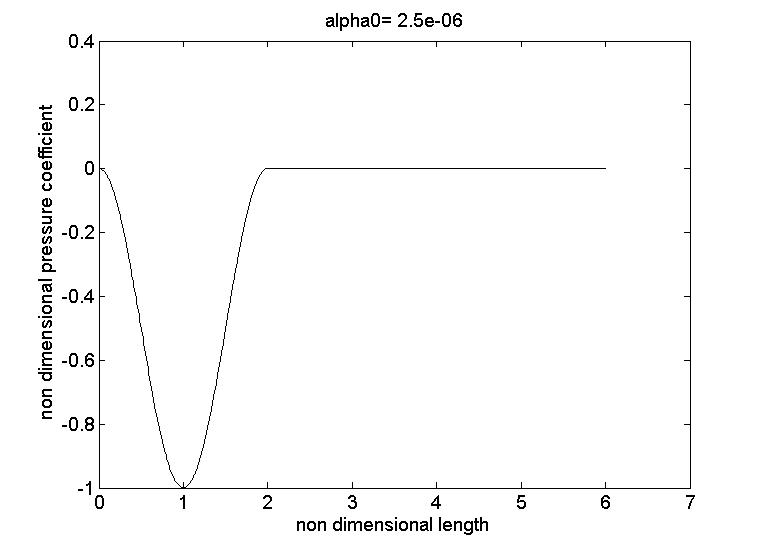}
\caption{Coefficient of Pressure plot for the case 1 of Liquid nitrogen}
\label{fig:chap5fig4}
\end{figure}

The void fraction plot, as shown in figure \ref{fig:chap5fig3}, increases continuously up to 3.8 $\times 10^{-4}$ after the nozzle exits from the location of 2.7 to the far downstream of the nozzle to reach the value of 1 which corresponds to fully vapour condition. From the plot for the coefficient of pressure shown in the figure \ref{fig:chap5fig4}, the same behaviour is inferred.\\

The magnitude of Pressure and Thermal source term is shown as follows, which shows that the magnitude of the thermal term for liquid nitrogen is 10 times the same for water, 
\begin{figure}
\centering
\includegraphics[scale=0.75]{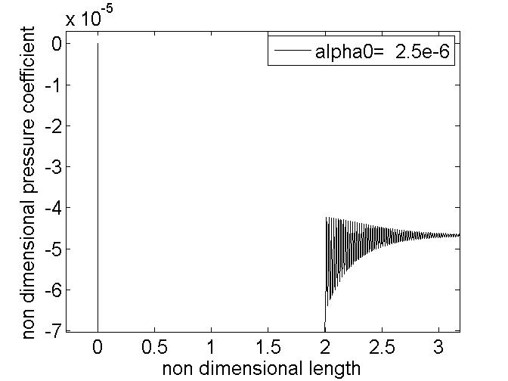}
\caption{Magnified view of coefficient of Pressure plot for case 1 of Liquid nitrogen}
\label{fig:chap5fig5}
\end{figure}
The magnitude of Pressure and Thermal source term is shown as follows, which shows that the magnitude of the thermal term for liquid nitrogen is 10 times the same for water, indicating the domination of thermal term in cryogenic liquids compared with isothermal fluids such as water  \ref{fig:chap5fig6}. The following figures  \ref{fig:chap5fig6}, \ref{fig:chap5fig7} shows the source terms plotted for case1

\begin{figure}[ht!]
\centering
\includegraphics[scale=0.5]{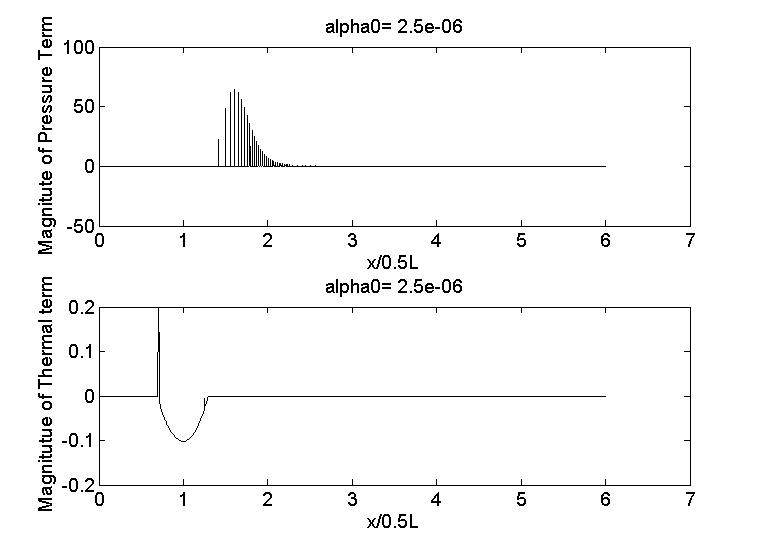}
\caption{Magnitude of Pressure and Thermal source term plot}
\label{fig:chap5fig6}
\end{figure}

\begin{figure}[ht!]
\centering
\includegraphics[scale=0.5]{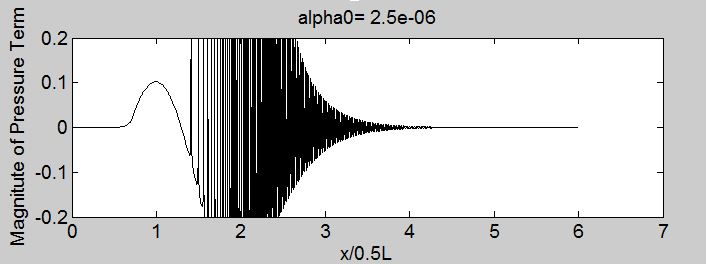}
\caption{Magnified view of pressure source term}
\label{fig:chap5fig7}
\end{figure}

\subsubsection{Case-2}
\paragraph{}

\indent Plot for the same are predicted for an Initial void fraction of $\alpha_{0} = 3.1 \times 10^{-6}$ and coefficient of pressure $ C_{p} $=0 \\

\begin{figure}[ht!]
\centering
\includegraphics[scale=0.4]{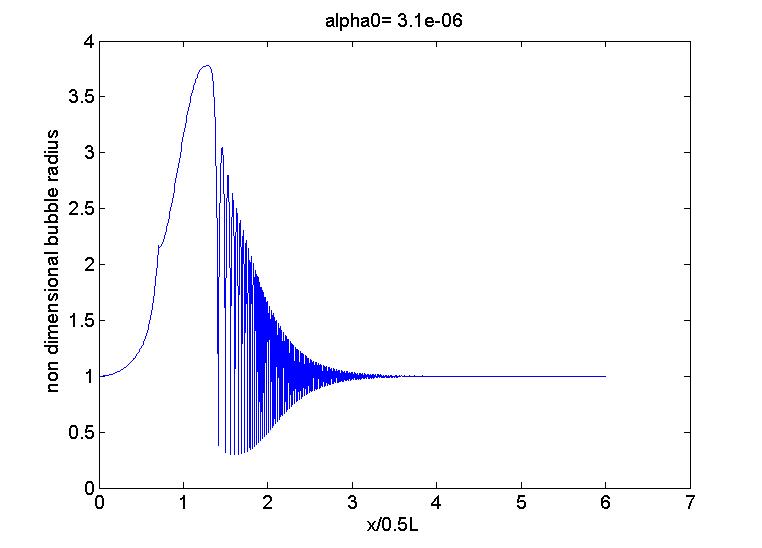}
\caption{Non-dimensional bubble radius plot for the case-2 of Liquid Nitrogen}
\label{fig:chap5fig8}
\end{figure}

\begin{figure}[ht!]
\centering
\includegraphics[scale=0.4]{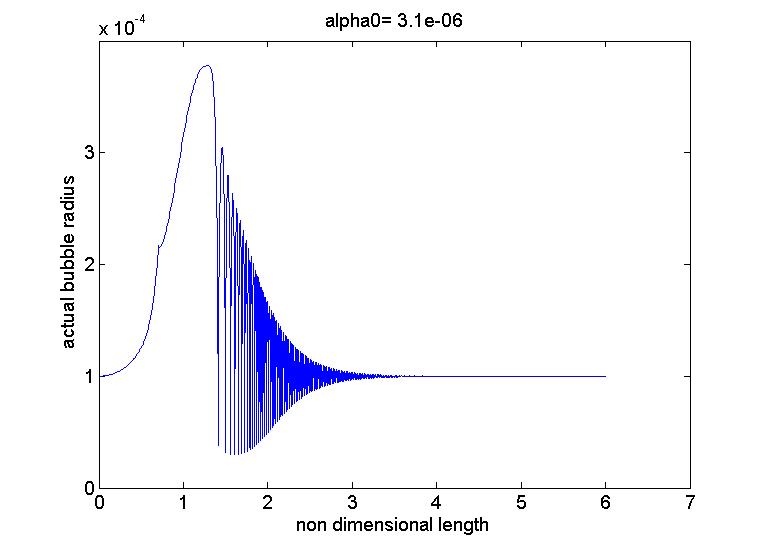}
\caption{Actual bubble radius plot for case 2 of Liquid Nitrogen}
\label{fig:chap5fig9}
\end{figure}

\begin{figure}[ht!]
\centering
\includegraphics[scale=0.45]{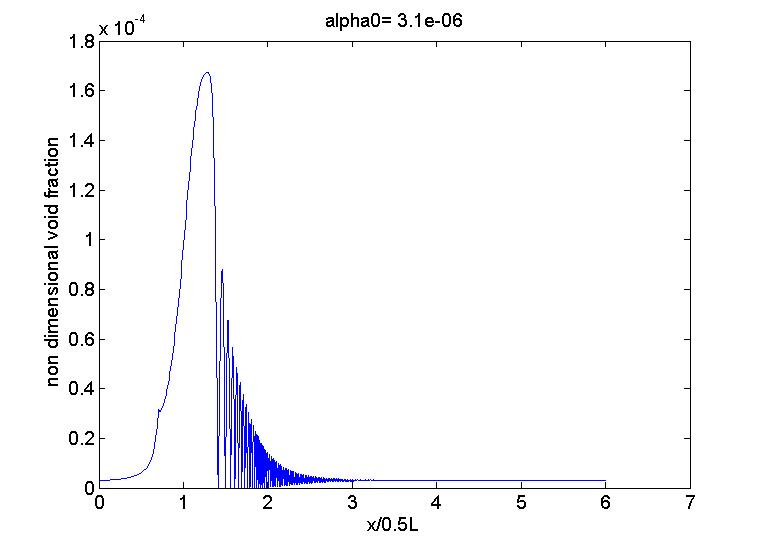}

\caption{Non-dimensional Void fraction plot for the case 2 of Liquid nitrogen}
\label{fig:chap5fig10}
\end{figure}

\begin{figure}[ht!]
\centering
\includegraphics[scale=0.4]{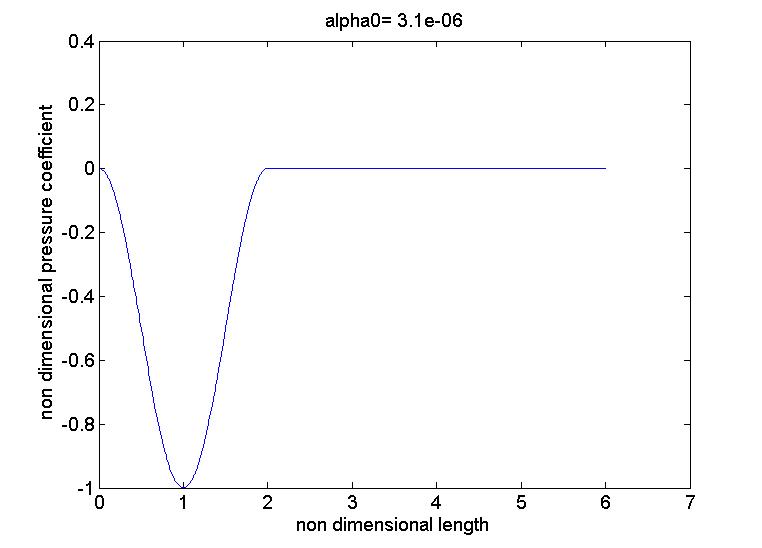}
\caption{Coefficient of Pressure plot for the case 2 of Liquid nitrogen}
\label{fig:chap5fig11}
\end{figure}

\begin{figure}[ht!]
\centering
\includegraphics[scale=0.4]{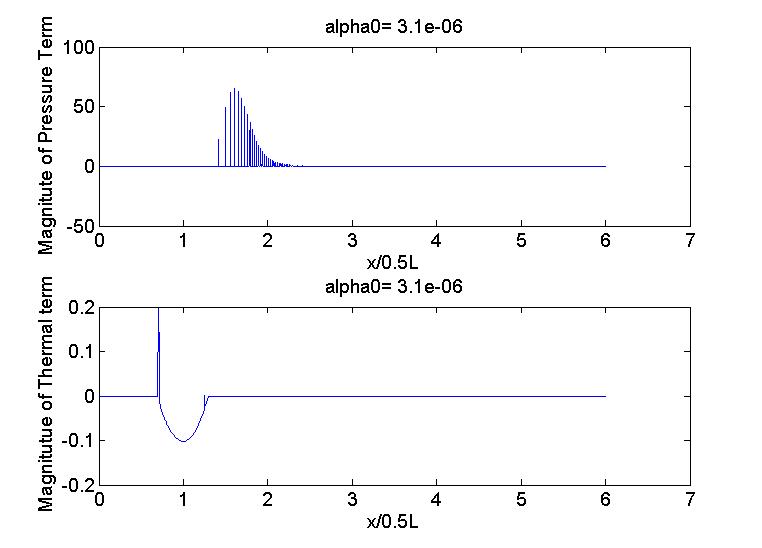}
\caption{Magnitude of Pressure and Thermal source term plot}
\label{fig:chap5fig12}
\end{figure}

\begin{figure}[ht!]
\centering
\includegraphics[scale=0.75]{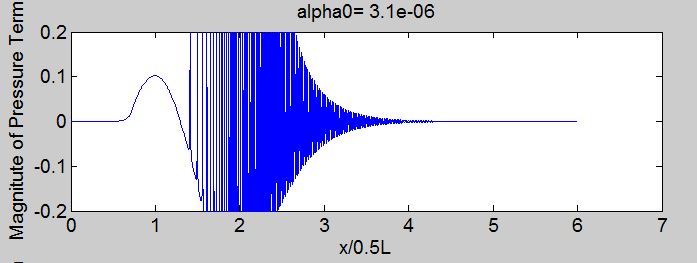}
\caption{Magnified view of pressure source term}
\label{fig:chap5fig13}
\end{figure}

\subsubsection{Case-3}
\paragraph{}
 Plot for the same is predicted for an Initial void fraction of  $\alpha_{0} = 3.0 \times 10^{-6}$ and coefficient of pressure $ C_{p} $ = 0
 
\begin{figure}[ht!]
\centering
\includegraphics[scale=0.45]{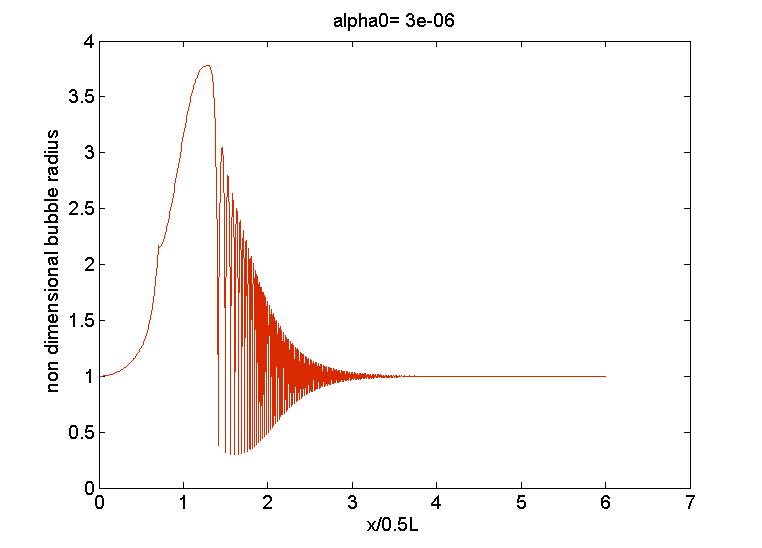}
\caption{Non-dimensional bubble radius plot for the case 3 of Liquid Nitrogen}
\label{fig:chap5fig14}
\end{figure}

\begin{figure}[ht!]
\centering
\includegraphics[scale=0.45]{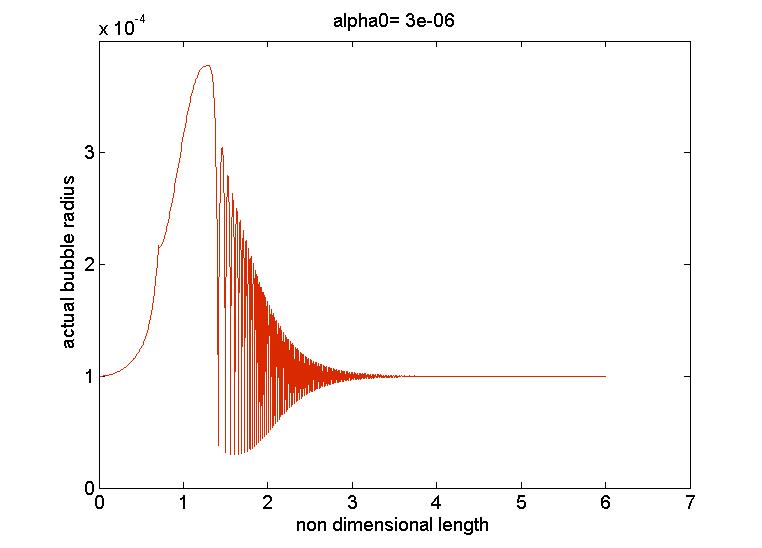}
\caption{Actual bubble radius plot for case 3 of Liquid Nitrogen}
\label{fig:chap5fig15}
\end{figure}

\begin{figure}[ht!]
\centering
\includegraphics[scale=0.45]{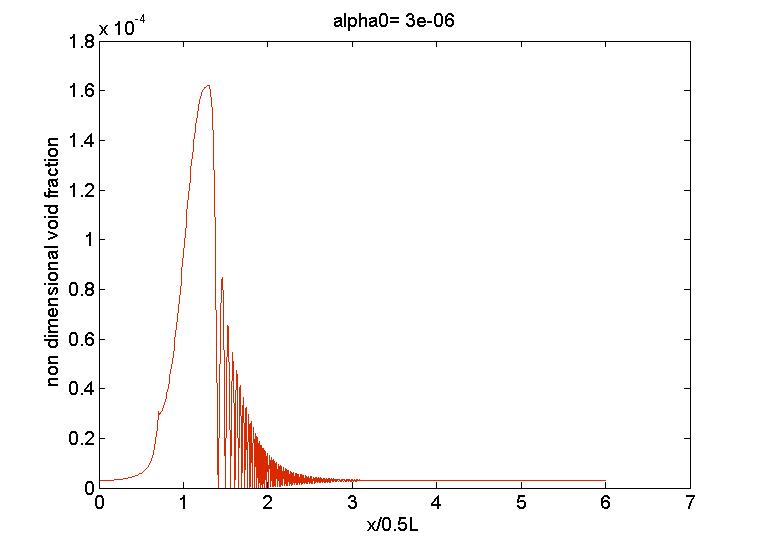}
\caption{Non-dimensional Void fraction plot for the case 3 of Liquid nitrogen}
\label{fig:chap5fig16}
\end{figure}

\begin{figure}[ht!]
\centering
\includegraphics[scale=0.45]{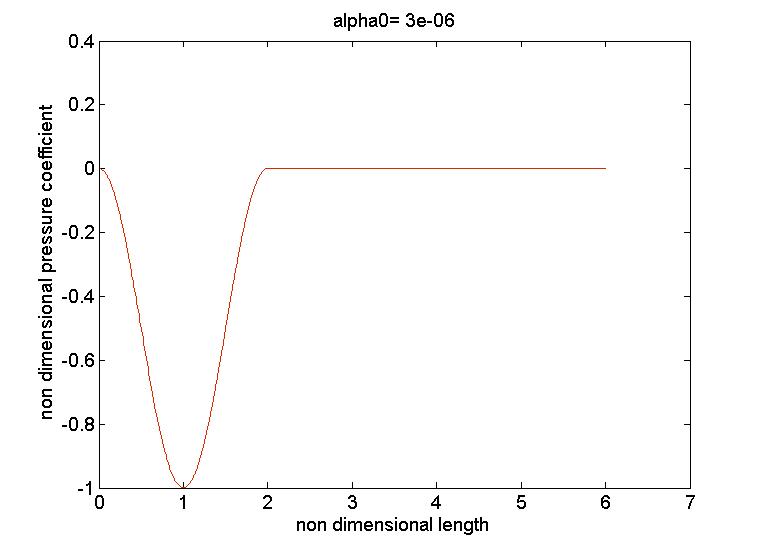}
\caption{Coefficient of Pressure plot for the case 3 of Liquid nitrogen}
\label{fig:chap5fig17}
\end{figure}

\begin{figure}[ht!]
\centering
\includegraphics[scale=0.4]{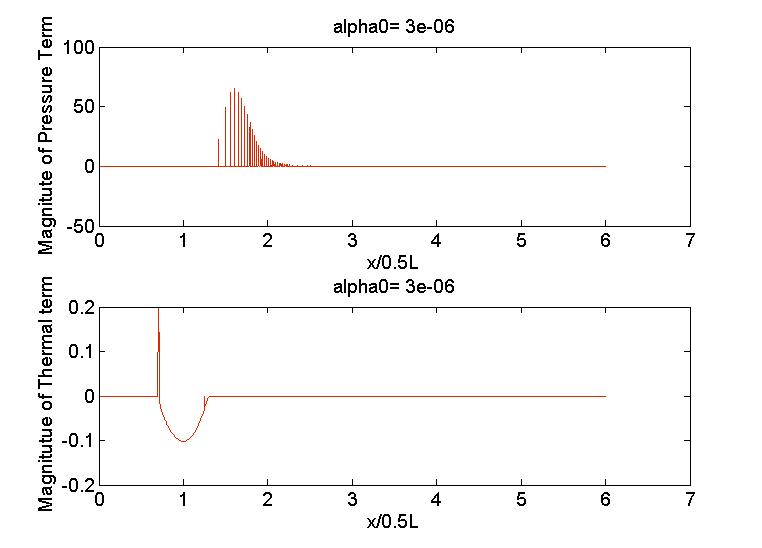}
\caption{Magnitude of Pressure and Thermal source term plot}
\label{fig:chap5fig18}
\end{figure}

The 1D numerical model for cavitating flows with the thermal effects resulted in successful benchmarking of the Wang $\&$ Brennen \cite{9} solution for water. The extension of the same model for cryogenic cavitating nozzle flows resulted in the damping solutions as the behaviour of cryogenic liquid nitrogen distinctly differs from the water. Overall, it can be ascertained that the 1D numerical model had limited predictability of the nature of cavitating flows, which are restricted to bubble radius, void fractions, coefficient of pressure, and the thermal and pressure source terms corresponding to a single bubble as the governing Rayleigh Plesset equation itself is valid for dynamics of the single bubble only. However, cavitation is a cluster phenomenon that demands the extension of the studies to two-dimensional numerical and experimental studies, whose attempted results are given in the subsequent sections.
 
\section{Two-dimensional model CFD Studies}
\paragraph{}
The following section systematically starts with an overview of the methodology adopted for 2D numerical studies using the commercial CFD solver ANSYS Fluent. The simulations have been carried out to capture the cavitating flows in a Venturi profile chosen to be symmetric, with a turbulence model to get insight into the actual experimental work. Simulations were done for cryogenic liquid nitrogen. The governing equations, such as continuity and momentum equations for the multiphase model, are solved using finite volume fluent code. Here, we discusses the viscous modeling, interfacial effects, solution methodology, and the boundary conditions used for the benchmarking test case of Hord et al. \cite{5} and the extension of the same for current 2D numerical work.
\subsection{Turbulence Modelling}
\paragraph{}

The actual cavitating nozzle and venturi flows are turbulent in nature.  Hence, a standard turbulence model has been incorporated into the simulations.  Realizable $k-\varepsilon$ was used in simulations. The chosen model of $k-\varepsilon$ is in line with the internal flows, and the coefficients for the same are the standard values built into the ANSYS-fluent. The model has two transport equations, one for turbulent kinetic energy k and another for turbulent dissipation $\varepsilon$. The two transport equations are given as follows
\begin{align}
\frac{\partial}{\partial t}(\rho k)+\frac{\partial}{\partial x_i}(\rho ku_i)=\frac{\partial}{\partial x_j} \bigg [ \bigg( \mu+\frac{\mu_t}{\sigma_k} \bigg ) \frac{\partial k}{\partial x_j}  \bigg ]+G_k+G_b - \rho \varepsilon - Y_m
\end{align}
And
\begin{multline}
\frac{\partial}{\partial t}(\rho \varepsilon)+\frac{\partial}{\partial x_i} (\rho \varepsilon u_i)=\frac{\partial}{\partial x_j} \bigg [ \bigg( \mu+\frac{\mu_t}{\sigma_{\varepsilon}} \bigg ) \frac{\partial \varepsilon}{\partial x_j}  \bigg ]+ \rho C_1 S_l \varepsilon+C_{1\varepsilon}\frac{\varepsilon}{k} ( G_k+ C_{3 \varepsilon b}G_b)
\\
- C_{2\varepsilon} \rho \frac{\varepsilon^2}{k+ \sqrt{\nu \varepsilon}} 
\end{multline}
Where,
C$_1$=max $\bigg [ 0.43+ \frac{\eta}{\eta +5} \bigg ]$ , $\eta=\frac{S_l k}{\varepsilon}$, $S_l=\sqrt{2S_{ij}S_{ij}}$ is the modulus of mean strain rate tensor. 
\\ \\

 The turbulent viscosity $\mu_t=\rho C_{\mu}\frac{k^2}{\varepsilon}$, where $C_{\mu}$ is a function of mean strain rate, the angular velocity of the system rotation, and the turbulence fields such as k and $\varepsilon$. The model constants are given by, 

$C_{1\varepsilon}=1.44,~~C_{2\varepsilon}=1.9,~~\sigma_k=1,~~ \sigma_{\varepsilon}=1.2$ 

The term G$_k$ indicates the production of turbulent kinetic energy and is given by, 
\begin{align*}
G_k=\mu_tS_l^2
\end{align*}

 The turbulence production due to buoyancy forces can be computed from the following relation
\begin{align*}
G_b=\beta_g \frac{\mu_t}{Pr_{t}}\frac{\partial T}{\partial x_i}
\end{align*}

\subsection{Modelling Interfacial effects}
\subsubsection{Schnerr Sauer model for Mass Transfer}

Estimation of mass transfer between the two phases has to be taken care of for complete cavitation modelling. The Schnerr and Sauer model addresses the net mass transfer from the liquid to the vapour. The vapour transport equation is given by
\begin{align*}
\frac{\partial}{\partial t} (\alpha \rho_{\nu})+ \nabla . (\alpha \rho_{\nu} \bar{\nu})=\frac{ \rho_{\nu}  \rho_{l}}{\rho} \frac{D \alpha}{Dt}
\end{align*}

The net source term for mass is given by the right side of the above equation as follows
\begin{align*}
\dot{m}=\frac{ \rho_{\nu}  \rho_{l}}{\rho} \frac{D \alpha}{Dt}
\end{align*}

Schnerr-Sauer model incorporated an expression that relates the vapour volume fraction to the
bubble number per unit volume of liquid, which is defined as
\begin{align*}
\alpha= \frac{\eta_b \frac{4}{3} \pi R_b^3 }{1+ \eta_b \frac{4}{3} \pi R_b^3}
\end{align*}

\indent R$_b$ is the bubble radius, and $\eta_b$ is the bubble population number per unit volume of the liquid.\\
This model assumes that the number of bubbles is neither created nor destroyed. To model bubble dynamics, the Rayleigh-Plesset equation in simple form was incorporated that was built with Schnerr-Sauer model, which is given by
\begin{align}
\frac{D \alpha}{Dt}= \sqrt{\frac{(P_{b}-P)^{2}}{\rho_{l}^{3}}}
\end{align}

\indent Where P$_b$ is the bubble pressure, P is the pressure, far away from the bubble. Using
the above relations, the mass source term is
\begin{align*}
\dot{m}=\frac{ \rho_{\nu}  \rho_{l}}{\rho} \alpha (1-\alpha) \frac{3}{R_b} \sqrt{\frac{(P_{b}-P)^{2}}{\rho_{l}^{3}}}
\end{align*}

\subsection{Interfacial Heat Transfer Effects}
\paragraph{}

\indent  Modelling the convective heat transfer coefficient $h_{b}$  was given by Ranz and Marshall \cite{14} model, which is inbuilt in Fluent and is given as follows
\begin{align*}
h_b=h_{pq}=\frac{6 k_{q} \alpha_{p} \alpha_{q} Nu_{p} }{d_p^2}
\end{align*}

\indent Here k$_q$ is the thermal conductivity of the q$^{th}$ face. The Nusselt number was determined from
\begin{align*}
Nu_{p}=2.0+0.6 Re_p^{1/2} Pr^{1/3}
\end{align*}

\indent Where Re$_p$ is the Reynolds number of the p$^{th}$ phase and Pr is the Prandtl number of the q$^{th}$ phase.

\subsection{Benchmark of  2D simulations with experimental work of Hord et al.}
\paragraph{}

\indent An attempt was made to benchmark the 2D numerical work with the experimental test case of Hord 121B \cite{5} for validating the cavitating length for liquid hydrogen flows in a venturi profile.

\indent A finite-volume mesh using Ansys workbench was performed for the actual venturi profile used by Hord. A grid independence study was done for the venturi, and a grid- independent solution was obtained for a mesh with 51,335 nodes and 50527 elements. The following figure shows the meshed model of the venturi profile.

\begin{figure}[ht!]
\centering
\includegraphics[scale=0.3]{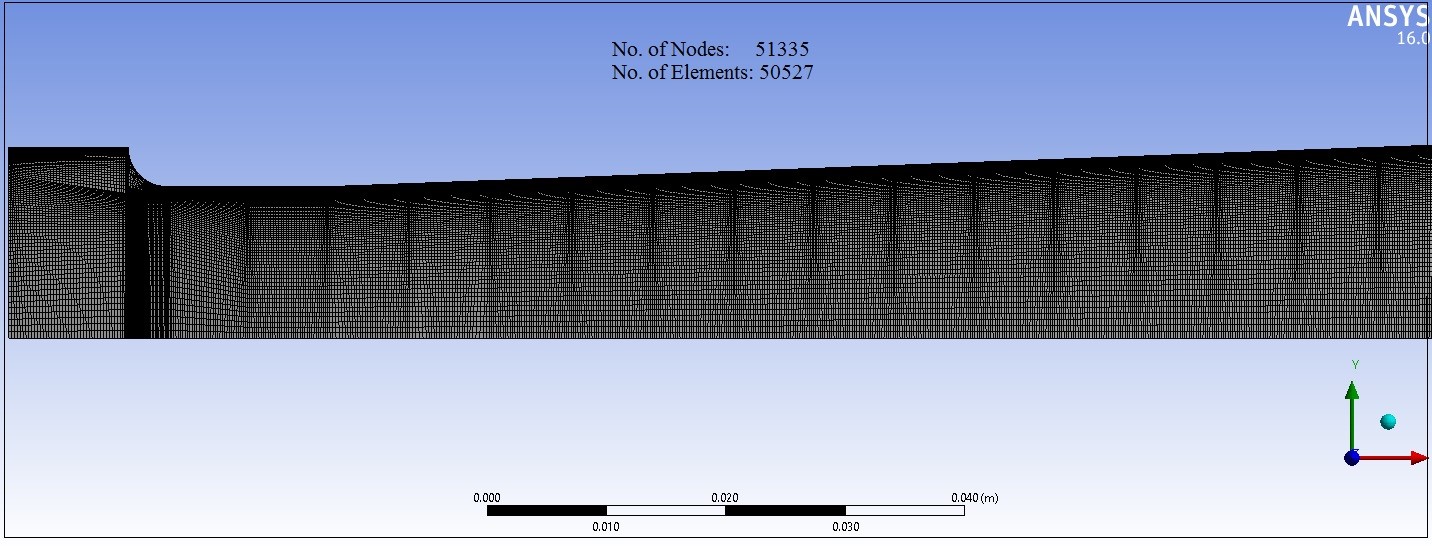}
\caption{Mesh details of Hord experimental venturi profile}
\label{fig:chp3fig11}
\end{figure}

\indent The simulation was done with the boundary conditions taken from the experimental data of Hord et al. \cite{5}.

\subsubsection{Solution Methodology}
\paragraph{}

\indent The simulations were performed using fluent with the Eulerian multiphase model. An implicit finite-volume scheme is based on a segregated pressure-based solver with SIMPLE as the pressure-velocity coupling. A second-order upwind scheme was used for continuity, momentum, energy, turbulent dissipation, and kinetic energy. 
\\ 
\indent Being turbulent flow, the wall y$+$ was assumed to be 30, and the resolution of grid element size near the boundary was 2 $\times$ 10$^{-04}$ m. A default standard wall function was chosen for solution methodology \cite{8}.
\\ 
\indent The residuals were set to 0.00001 for continuity, velocity, K, and $\varepsilon $. For the energy equation and volume fraction, the residuals were set to be 1$^{-06}$.
\\ 
\indent Following are the simulation results for the contour of Vapour fraction obtained using Ansys fluent

\subsubsection{Simulation Results}
\paragraph{}

\begin{figure}[ht!]
\centering
\includegraphics[scale=0.75]{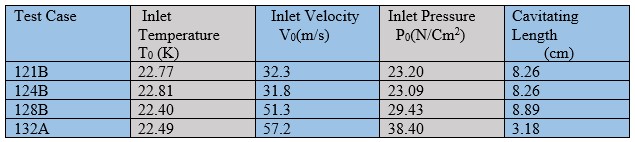}
\caption{Table showing experimental test cases of Hord}
\label{fig:b}
\end{figure}

\indent The following table\ref{fig:b} is the actual experimental result of Hord for various test cases from the flow visualization study.

\indent The following figure\ref{fig:chp3fig12} shows the simulation results of Hord test case 121B. The blue fill indicates the complete liquid phase, and the red fill indicates the complete vapour phase.
\begin{figure}[ht!]
\centering
\includegraphics[scale=0.6]{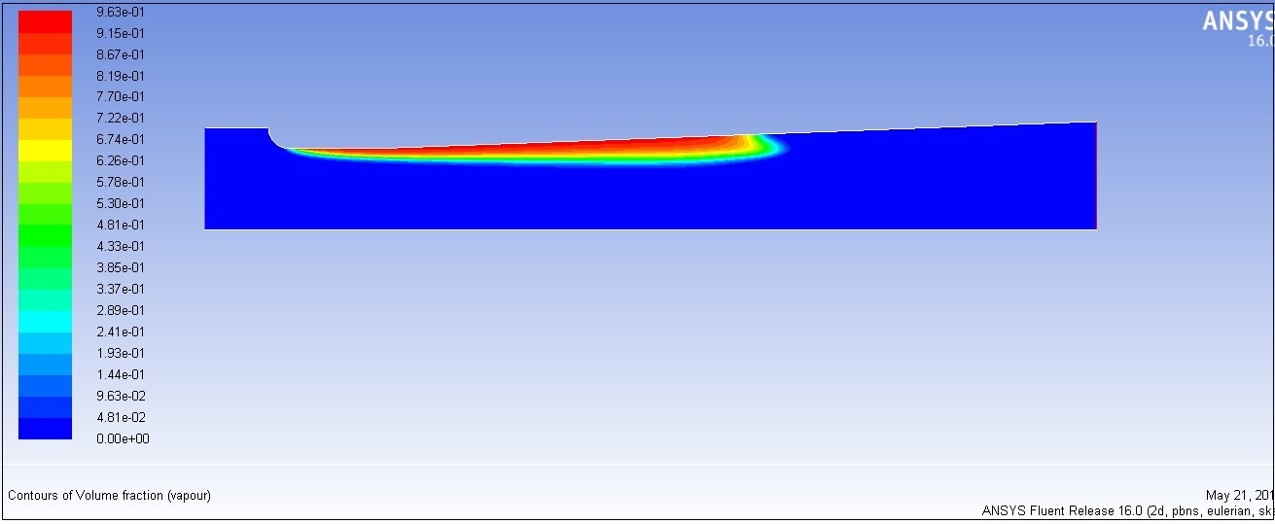}
\caption{Contour of volume fraction from the simulation results of Hord profile}
\label{fig:chp3fig12}
\end{figure}

\begin{figure}[ht!]
\centering
\includegraphics[scale=0.5]{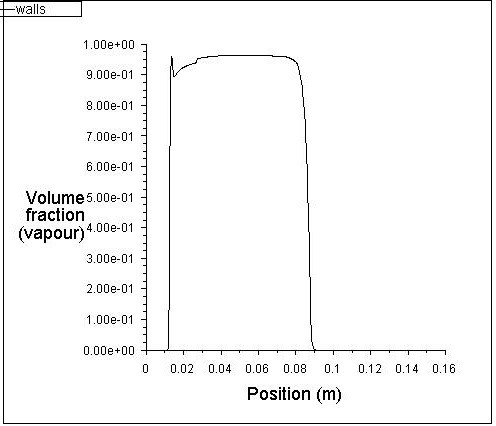}
\caption{Plot for volume fraction as a function profile length}
\label{fig:chp3fig13}
\end{figure}
\indent From the vapour volume fraction plot\ref{fig:chp3fig13}, it is evident that the vapour phase dominates in the cavitating regime, and the plot shows the cavitating length of 0.085 m or 8.5 cm.

The simulation results for other test cases are as follows:

\begin{figure}[ht!]
\centering
\includegraphics[scale=0.6]{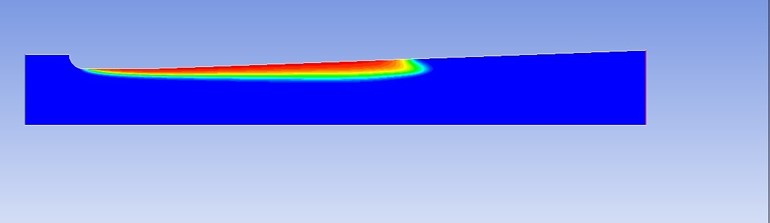}
\caption{Contour of volume fraction for test case 124B}
\label{fig:chp3fig14}
\end{figure}

\begin{figure}[ht!]
\centering
\includegraphics[scale=0.6]{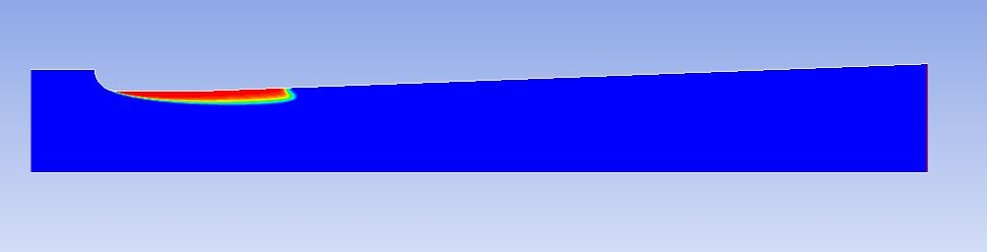}
\caption{Contour of volume fraction for test case 126A}
\label{fig:chp3fig15}
\end{figure}

\begin{figure}[ht!]
\centering
\includegraphics[scale=0.6]{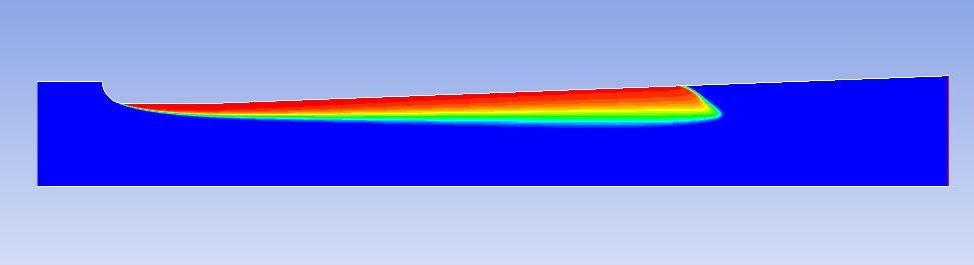}
\caption{Contour of volume fraction for test case 128A}
\label{fig:chp3fig16}
\end{figure}

\subsubsection{Estimating Cavitating length }
\paragraph{}

\begin{figure}[ht!]
\centering
\includegraphics[scale=0.6]{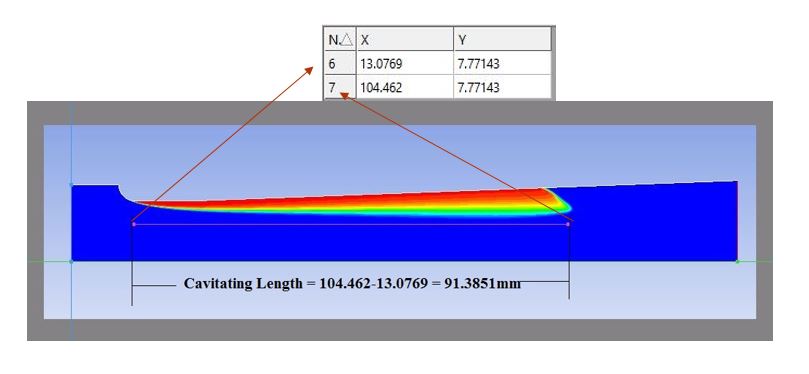}
\caption{Cavitation length measurement details}
\label{fig:chp3fig17}
\end{figure}

\indent The details of the measured Cavitating length from the contour of the void fraction is shown in the below figure\ref{fig:chp3fig17}

In the above figure, the cavitating length indicates the domination of vapour phase.
The simulation result for a sample case, as shown in the figure\ref{fig:chp3fig17} indicated a Cavitating Length of 9.13 cm against the actual 8.89 cm of experimental cavitating length for test case 128B measured by Hord et al. \cite{5}. Exact matching of simulation and experimental data was impossible as the solutions are independent of numerical solution methodology and wall treatments. Still, the computation results closely match the experimental result of Hord et al. \cite{5}. as the error in prediction is less than 4$\%$. 

\begin{figure}[ht!]
\centering
\includegraphics[scale=0.75]{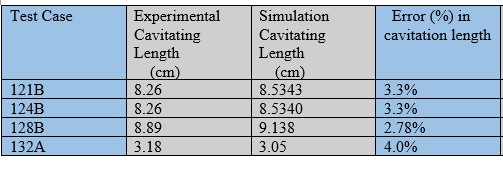}
\caption{Table showing error in the simulation results of Hord}
\label{fig:chp3fig18}
\end{figure}

Error prediction in cavitating length against the actual experimental length for different test cases are shown in the figure\ref{fig:chp3fig18} containing the table
Similar work was done by Rodio et al. \cite{8}, and the contour of the void fraction of the vapour obtained for the Hord test case 121B \cite{5} is shown in the following figure.

\begin{figure}[ht!]
\centering
\includegraphics[scale=1]{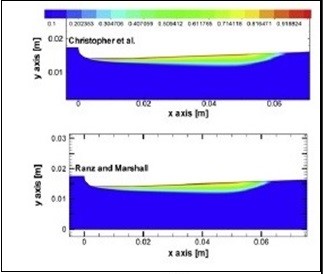}
\caption{Contours of the void fraction obtained by Rodio et.al \cite{8} for Hord test case 121B \cite{5}}
\label{fig:chp3fig19}
\end{figure}

It was observed that the simulation results by Rodio et al. \cite{8} overpredicts the experimental cavitating length observed by Hord et al. \cite{5}. The following measurement proofs validate the above interpretation of the result.

\begin{figure}[ht!]
\centering
\includegraphics[scale=0.8]{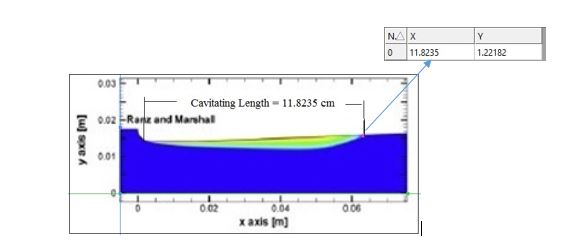}
\caption{Cavitation Length measurement details for Rodio et.al \cite{8}  work}
\label{fig:chp3fig20}
\end{figure}

The calculated cavitating length from the above contours of volume fraction gives a length of 11.8235 cm compared to the actual experimental cavitating length of 8.26 cm, measured by Hord et al. \cite{5}.
However, the wrong prediction of cavitation length may not be precisely due to the model used, as mesh methods and chosen wall functions have significant influences on the  convergence of numerical solution.
Having benchmarked the 2D numerical model for cryogenic cavitating venturi with Hord et al. \cite{5} experimental test cases for predicting the cavitating length in the venturi profile. The same is now extended for the current work, the details of which are given in the following section.
\section{2D simulation for Current Work}
\subsection{Computational Model Details}
\paragraph{}

\indent A steady-state, two-dimensional, incompressible, turbulent Eulerian mixture model was developed based on finite-volume formulation. The details of the computational model are shown below
\begin{figure}[ht!]
 \begin{center}
  \includegraphics[scale=0.75]{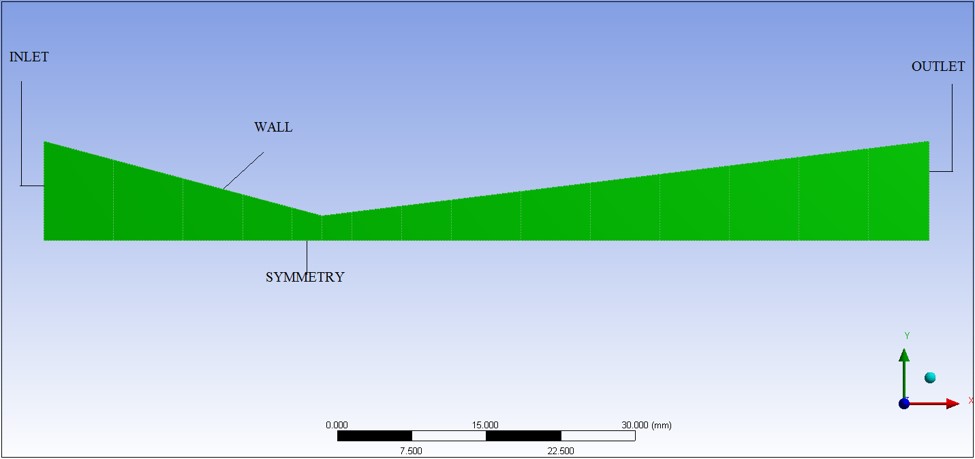}
  \caption{Computation model with domain or boundary}
  \label{fig:chp3fig21}
 \end{center}
\end{figure}

\indent The figure\ref{fig:chp3fig21}depicts the actual 2D model of the venturi region of the test section developed for simulations. The dimensions correspond to the actual dimensions of the experimental setup. The boundaries include an inlet, wall, symmetry, and the outlet.

\indent Domain discretization or meshing was done with Ansys and Pointwise. The entire domain was discretized using structural quadrilateral elements. The wall y$^{+}$ requirement for $k - \varepsilon$ model was assumed to be around 30 and accordingly the thickness of the first computational grid was fixed.

\subsection{Boundary conditions}
\paragraph{}
The entire domain was divided into three regions: inlet, outlet, and walls. Only the inlet and outlet boundary conditions were found to be very important as they play a very significant
role in the convergence of solutions.

 The inlet and outlet pressure conditions from the experiments were used as  boundary conditions, with the inlet temperature assumed as
 77 K for the liquid phase of liquid nitrogen. Also, the inlet vapour fraction is given 0 as the inlet condition is fully liquid.

 Adiabatic wall boundary conditions were given as there were no assumptions of heat transfer effects between the wall boundary and fluid.

 The outlet boundary condition was atmospheric pressure, as the
experiment's venturi profile opens to the atmosphere. Liquid temperature condition at the outlet is assumed to be 77 K, and vapour fraction at the outlet is assumed to be 0 indicating a single-phase liquid.

\subsection{Solution Methodology}
\paragraph{}

 The same solution methodology was used for the Hord benchmark simulation. It is the exact segregated pressure-based solver with SIMPLE as the pressure-velocity coupling. A second-order upwind scheme was used for continuity, momentum, energy, turbulent dissipation, and kinetic energy.

 under-relaxation parameters for vapourization mass were set to 0.5, and the volume fraction was set to 0.95.

 The residuals were set to be 0.001 for continuity, velocity, K, and $\varepsilon$. For Energy and volume fraction, the residuals were set to be 1 e$^{-06}$. The simulation results for the contour of vapour fraction obtained using Ansys fluent are discussed in following section.

\subsection{Grid Independence}
\paragraph{}

\indent A grid independence study was carried out to find the optimum number of grid points that ensures correct capturing of flow characteristics and ensures that the solution does not change with further increase in the number of grid points.

 The domain was discretized using finite volume meshing in Ansys with grid elements of 8346, 40926, and 102482 elements. It was found that a grid-independent solution was obtained with a mesh of 40926 grid elements. The following picture shows the meshing details. For near-wall treatment, a default wall y$^{+}$ $>$ 30 for turbulent flow was assumed, and accordingly, the size of the first grid element near the wall was chosen for meshing. The following picture shows the meshing details

\begin{figure}[ht!]
 \begin{center}
  \includegraphics[scale=0.75]{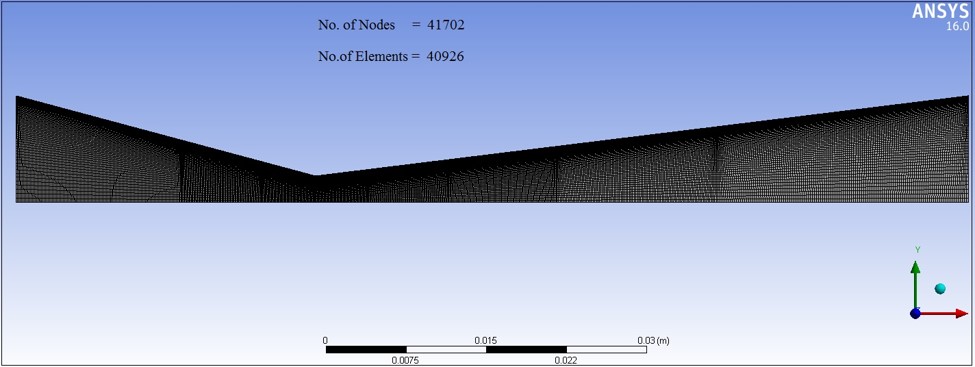}
  \caption{Mesh detail for the venturi}
  \label{fig:chp3fig22}
 \end{center}
\end{figure}

\subsection{2D Numerical results}
\paragraph{}
 This section discusses the simulation results of the symmetric venturi corresponding to 40926 grid elements. The Boundary conditions used in the simulations were in line with experimental observations.

\subsubsection{Boundary Conditions}
\begin{figure}[ht!]
\centering
\includegraphics[scale=0.8]{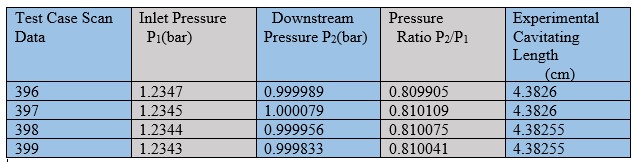}
\caption{Table showing experimental data}
\label{fig:chap5fig19}
\end{figure}

\paragraph{}
The following table \ref{fig:chap5fig19} gives the details of the experimental observations. The same values of pressure are used for 2D simulations as boundary conditions.

 The simulation results obtained from Ansys-Fluent for various test cases are as follows.

\subsubsection{Contour of Volume fraction(Liquid)}
\paragraph{}

\begin{figure}[ht!]
\centering
\includegraphics[scale=0.6]{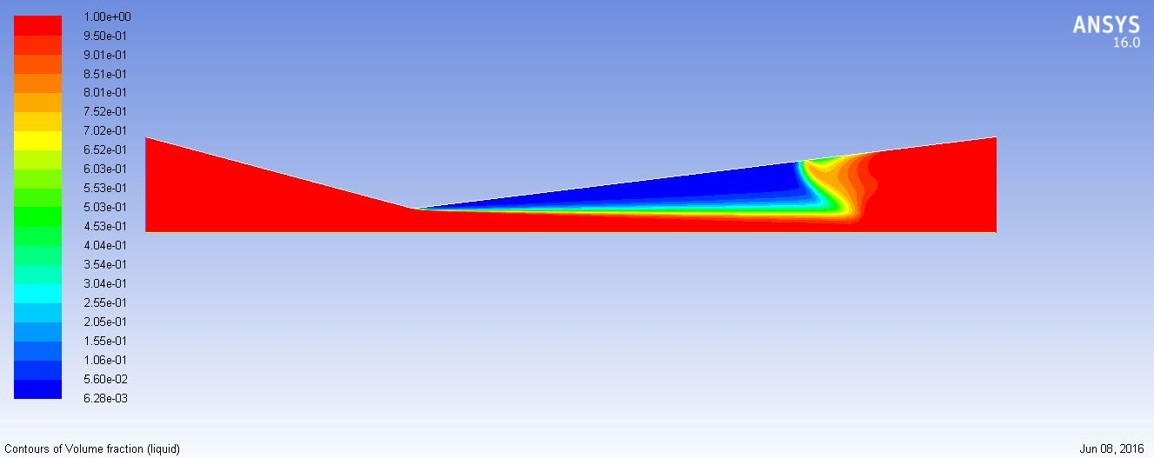}
\caption{Contours of volume fraction for test case 396}
\label{fig:chap5fig20}
\end{figure}

\begin{figure}[ht!]
\centering
\includegraphics[scale=0.6]{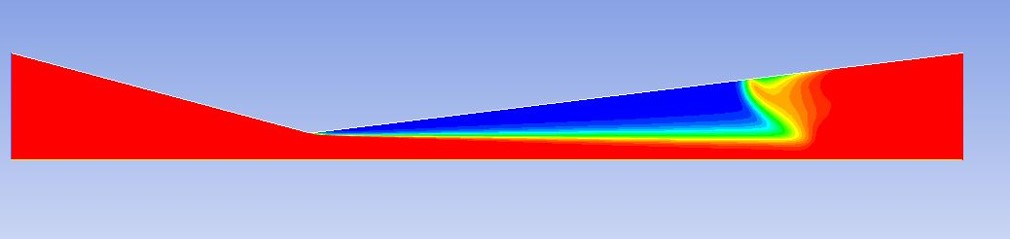}
\caption{Contours of volume fraction for test case 397}
\label{fig:chap5fig21}
\end{figure}

\begin{figure}[ht!]
\centering
\includegraphics[scale=0.6]{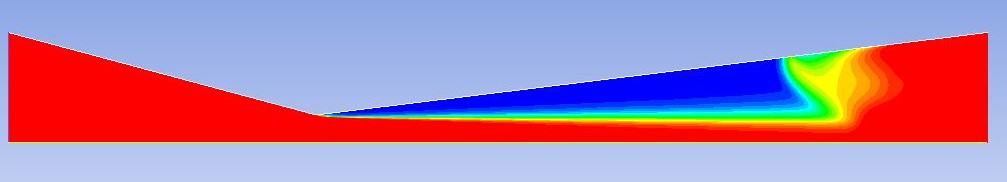}
\caption{Contours of volume fraction for test case 398}
\label{fig:chap5fig22}
\end{figure}

\begin{figure}[ht!]
\centering
\includegraphics[scale=0.6]{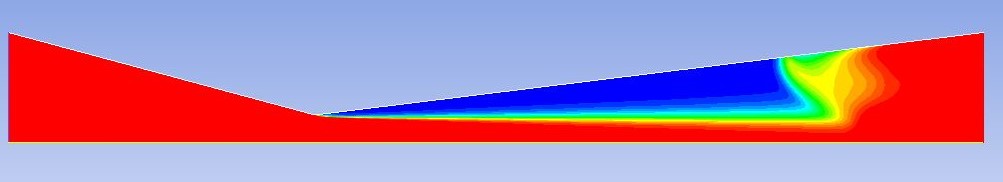}
\caption{Contours of volume fraction for test case 399}
\label{fig:chap5fig23}
\end{figure}

\indent  The blue fill indicates a complete vapour phase, and the red fill indicates a fully liquid phase. The yellow region corresponds to a vapour fraction of 0.6, which indicates a thin two-phase regime, which is a point where the vapour is ready to convert back to the single-phase liquid. It is evident from the above figures that there is a distinct cavitation regime from the throat section up to a certain length in the downstream section of the venturi that corresponds to the cavitating length of the venturi.

\subsubsection{Cavitating Length Measurement}
\paragraph{}

 The sample calculation of the cavitating length is shown as follows

\begin{figure}[ht!]
\centering
\includegraphics[scale=0.6]{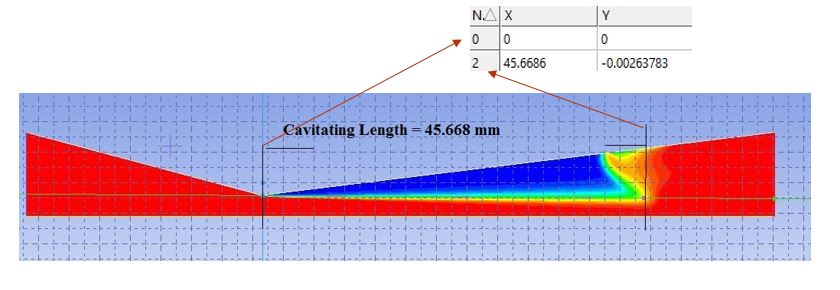}
\caption{Sample cavitating length measurement for symmetric venturi}
\label{fig:chap5fig24}
\end{figure}

\begin{figure}[ht!]
\centering
\includegraphics[scale=0.9]{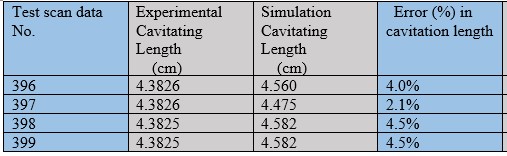}
\caption{Table showing the error in simulation cavitating length}
\label{fig:chap5fig26}
\end{figure}

 From the above calculation, the cavitating length was found by subtracting the x-coordinate at the end of the vapour regime from the x-coordinate at the throat location, which results in the cavitating length of  45.668 mm or 4.56 cm. 
\indent A similar calculation is done for other cases, and the error in the cavitating length of simulation against the experiments is shown in the following table \ref{fig:chap5fig26}

It is concluded that the above 2D numerical simulation model for cavitating venturi flows resulted in the reasonable benchmark of the experimental test case of Hord et al. \cite{5} and gave an insight for the prediction of cavitating length. The extension of the same for the current symmetric venturi resulted in the prediction of void fraction and the cavitating length, which resulted in an error of less than 4.5$\%$, which is a reasonable comparison of the cavitation length with the actual experimental flow visualization study. 
\FloatBarrier
\section{Experimental Observations}
\paragraph{}
 This section puts forth the experimental observations of two experiments, one with the acrylic venturi test section and the other with the aluminium venturi test section. The acrylic venturi was incompatible with the cryogenic temperature flow conditions, resulting in the crack of the test section itself.
  \begin{figure}[ht!]
\centering
\includegraphics[scale=0.5]{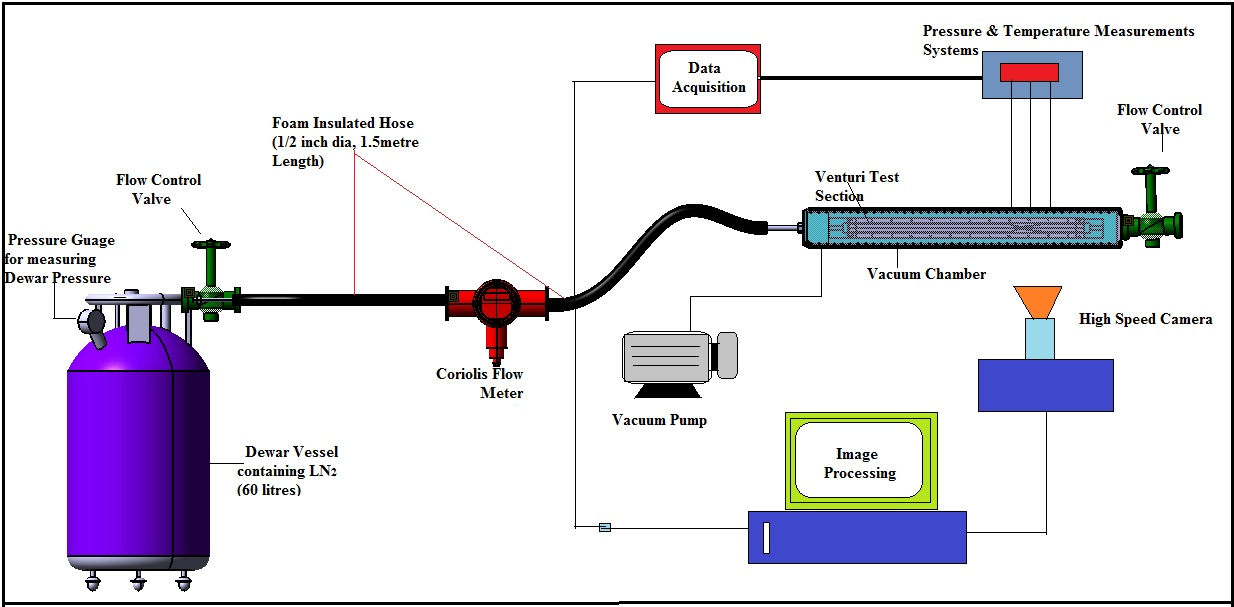}
\caption{Schematic of the Experimental setup.}
\label{fig:expt_setup_1}
\end{figure}
 \begin{figure}[ht!]
\centering
\includegraphics[scale=2]{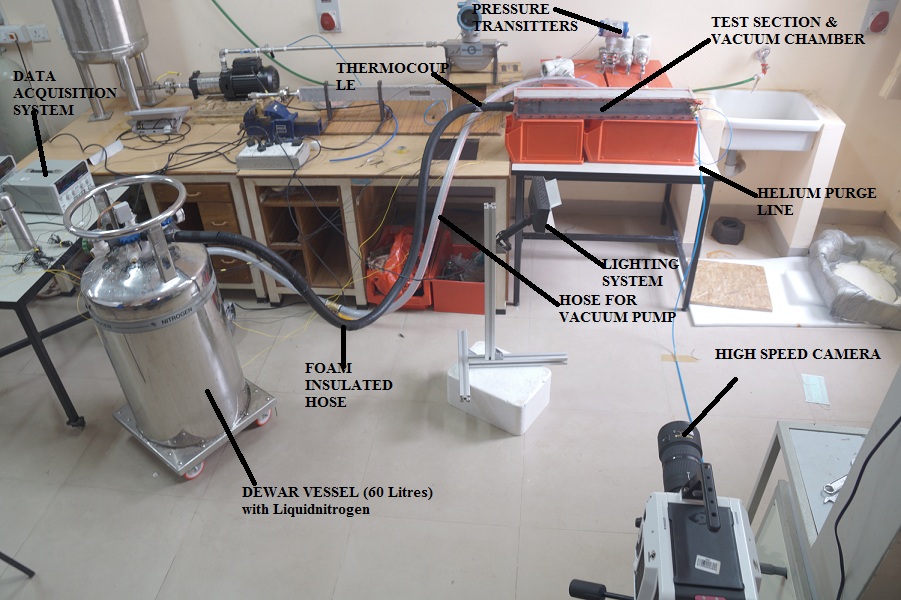}
\caption{Experimental setup.}
\label{fig:expt_setup_2}
\end{figure}
Following this, an attempt was made using the same experiment with the aluminium venturi. The flow visualization images during the cavitating time of venturi are presented, and an attempt was made to evaluate the experimental cavitation length from these visualization images.
\subsection{Observations from experiment - 1}
 The experiment was carried out with an initial Dewar pressure of 2.5 bar, and the outlet of the venturi test section was opened to atmospheric pressure. After an initial transient time for chilling down of the set-up, the developed flow in the test section appeared 4.30 mins from the start of the experiment. Immediately, the nature of the flow of liquid nitrogen turned stratified while flowing inside the test section. This was because of the crack, which developed in the adapter region upstream of the test section, which had a metal-to-acrylic contact. Also, it was observed that the acrylic test section, in due time, resulted in multiple cracks at different spots, which can be seen from the following figure, which was taken from the 42547$^{th}$ frame record of the high-speed camera.
\begin{figure}[ht!]
\centering
\includegraphics[scale=0.6]{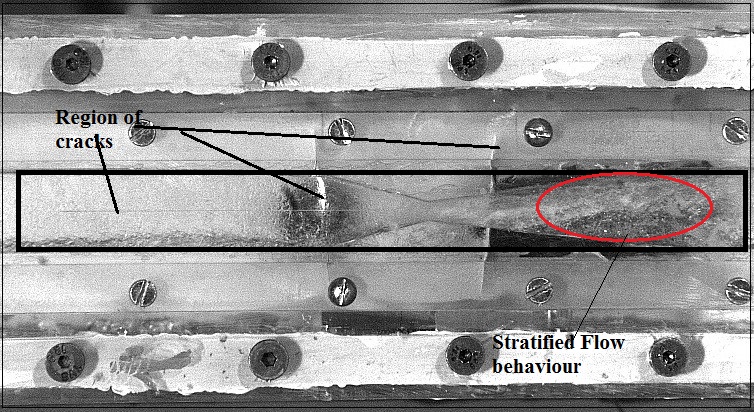}
\caption{Test section of the experiment - 1}
\label{fig:chap5fig27}
\end{figure}

In the above experiment, there was no sign of any developed single-phase flow of liquid nitrogen, and there was no recognized cavitating regime because of the stratified nature of the flow, which was believed to be caused by the cracks in the test section. So, the image data obtained did not give any physical insight into the cavitating nature of the flow and the length. However, the pressure measurement readings closely matched the cavitating regime's operating conditions.

\subsection{Observations from experiment - 2}
The experiment was carried out with an initial Dewar pressure of 1.2 bar, and the outlet of the aluminium venturi test section was opened to the atmosphere. The time taken for chilling down was reduced compared to the previous experiment, as the entire test section was wound with foam insulation. The developed flow in the test section appeared at 2 min from the start of the experiment. The image was captured at 198$^{th}$ sample scan of the data acquisition system from the video taken by the high-speed camera.

\begin{figure}[ht!]
\centering
\includegraphics[scale=0.5]{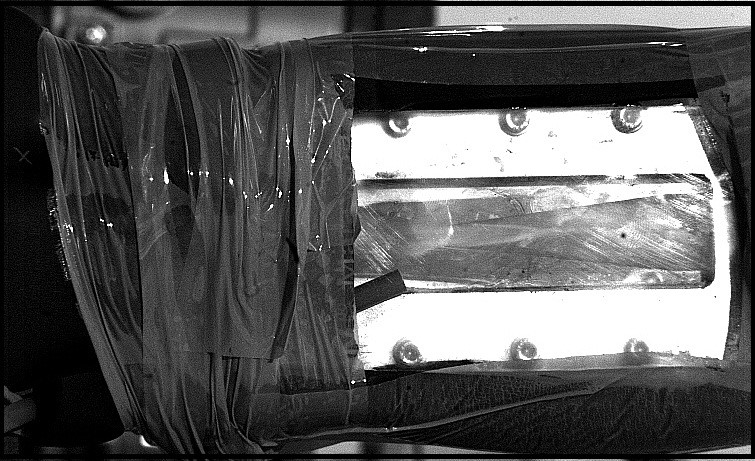}
\caption{Flow visualization image captured during experiment - 2}
\label{fig:chap5fig28}
\end{figure}

The figure \ref{fig:chap5fig28} shows a distinct vapour regime, which starts from the throat region and extends to some regions in the diverging section of the venturi. The reading from the pressure measurement indicated a near-cavitation operating condition. 

\indent The following figure \ref{fig:chap5fig30} shows the images and corresponding cavitating length for the test cases.
\begin{figure}[ht!]
\centering
\includegraphics[scale=1]{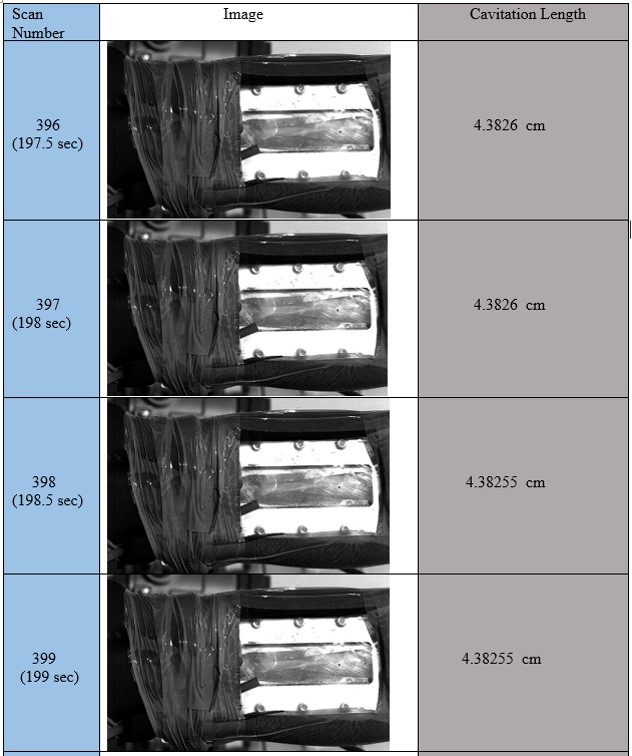}
\caption{Figure showing various test cases}
\label{fig:chap5fig30}
\end{figure}
\FloatBarrier
\section{Conclusion}
The attempted numerical 1D model based on the extension of the Wang and Brennen [9] model with
an additional thermal term in line with the numerical 1D model of Rodio [8] was implemented using a finite-difference code in MATLAB. It was inferred that the thermal term in a cavitating water flow had a negligible role to play in bubble growth or collapse. On the other hand, this thermal term had a significant influence on the bubble radius for the liquid nitrogen case.
The 2D numerical model was implemented in ANSYS Fluent, and the following are the outcomes
of the simulation studies.\\
• The cavitating length predicted and compared with the experimental results of Hord et al. [5] for validating the developed 2D numerical model.\\
• A similar procedure was extended for the chosen cavitating venturi using the experimental data as the boundary conditions, and the cavitating length was predicted for the same.\\
The following are the outcomes from the experimental observations:\\
• The acrylic venturi test section did not withstand the low-temperature liquid nitrogen flow,
resulting in leaks near the upstream venturi and altering the flow pattern to stratified
conditions. Hence, a distinct cavitating flow could not be observed in
experiment 1.\\
• The experiment-2, with an aluminum venturi, showed distinct cavitating regimes
and the cavitating lengths were estimated from the recorded high-speed camera experimental images.\\

\textbf{Acknowledgement}
We acknowledge Prof.Kannan Iyer and Dr.Nandhakumar (LPSC) for their active discussions. Mr.Dinesh and Mr.Bipin of Thermal Engineering lab are recognized for their assistance with Experiments.


\addcontentsline{toc}{chapter}{REFERENCES}

\textbf{Appendix A:  Numerical Method for Modelling 1D Cavitating nozzle / Venturi flow using Matlab}

\paragraph{}

\begin{figure}[ht!]
\centering
\includegraphics[scale=0.8]{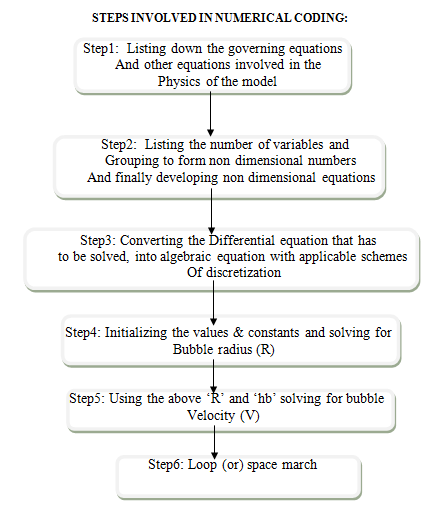}
\caption{steps involved in numerical coding}
\label{fig:chp3fig1}
\end{figure}

\textit{\textbf{Step1: List of Governing Equations and closures:}}
\\ 
Continuity and Momentum equation
\begin{center}
$\dfrac{\partial(1-\alpha)A}{\partial t}+\dfrac{\partial(1-\alpha)Au}{\partial x}=0,$ 
\\
$\dfrac{\partial u}{\partial t}+u\dfrac{\partial u}{\partial x}=-\dfrac{1}{2(1-\alpha}\dfrac{\partial C_{p}}{\partial x}$
\end{center}

Pressure Coefficient
\begin{center}
$C_{p}(x,t)=\dfrac{P(x,t)-P_{0}}{0.5\rho_{l}u_{0}^{2}}$
\end{center}

Void fraction equation
\begin{center}
$ \alpha(x,t)=\dfrac{\dfrac{4}{3}\pi \eta R(x,t)^{3}}{\left( 1+\dfrac{4}{3}\pi \eta R(x,t)^{3} \right)  } $
\end{center}

Modified Rayleigh-Plesset equation
\begin{center} 
$\left[ R \ddot{R} + \dfrac{3}{2} \dot{R}^{2} \right] +\dfrac{2S}{\rho_{L}R} +\dfrac{4\mu}{\rho_{L}}\dfrac{\dot{R}}{R} =  \dfrac{P_{g0}}{\rho_{L}}\left[ \dfrac{R_{0}}{R} \right]^{3\gamma} + \dfrac{\left[ P_{\nu}(T_{\infty})-P_{\infty}(t) \right]}{\rho_{L}} - \dfrac{dP_{\nu}}{dT}\dfrac{\rho_{\nu} L}{\rho_{L}}\dfrac{\dot{R}}{h}$
\end{center}

Ranz $\&$ Marshall convective heat transfer model
\begin{center}
$h_{b}=\dfrac{Nu_{b} K_{l}}{2R}$ \\ 
$ Nu_{b}=2+0.6Re_{b}^{\frac{1}{2}}Pr^{\frac{1}{3}} $ \\ 
$ Re_{b} =2R\vert \nu-u \vert/\lambda_{l} $\\ 
$ Pr=\dfrac{C_{pl}\mu_{l}}{K_{l}} $
\end{center}

Bubble momentum equation
\begin{center}
$\rho_{\nu}\dfrac{D\nu}{Dt}+\dfrac{1}{2}\rho_{l}\left( \dfrac{D\nu}{Dt}-\dfrac{Du}{Dt} \right) = -\dfrac{\partial P(x,t)}{\partial x}-\dfrac{3}{8}\rho_{l}C_{D}\times \dfrac{(\nu-u)\vert\nu-u\vert}{R}$
\end{center}

\textit{\textbf{Step2:  Non-dimensionalizing the Equations:} }                      
\\ \\
\indent All the equations mentioned above were non-dimensionalized by using following dimensionless variables  
\begin{center}
$\overline{u} = \dfrac{u}{u_{0}} ; \overline{x} = \dfrac{x}{R_{0}} ; \overline{R} = \dfrac{R}{R_{0}} ; \overline{t} = \dfrac{tu_{0}}{R_{0}} ; \overline{v} = \dfrac{v}{u_{0}} ; \overline{\eta} = \dfrac{\eta}{R_{0}^{3}} ; \overline{L} = \dfrac{L}{R_{0}} $
\end{center}

\indent where the subscript "0" corresponds to the upstream value. Some of the non-dimensional numbers, such as
\begin{center}
$ Re=\dfrac{\rho_{L}uR}{\mu_{E}}  ; We=\dfrac{\rho_{L}uR}{S} ; \sigma = \dfrac{P - P_{v} }{\dfrac{1}{2} \rho_{L} U^{2} }  $
\end{center}

\indent After applying these dimensionless terms to the actual steady case equation, the final system of dimensionless equations is of the form
\begin{center}
$\left( 1- \alpha\right) \overline{u} \overline{A} =  \left( 1- \alpha_{0} \right) = const $ 
\end{center}
\begin{center}
$\overline{u} \dfrac{d \overline{u}}{d \overline{x}}  = - \dfrac{1}{2 \left( 1- \alpha\right)} \dfrac{dc_{p}}{d \overline{x}}$
\end{center}
\begin{multline}
\overline{R}  \left( \overline{u}^{2}  \dfrac{d^{2}R}{dx^{2}} + \overline{u} \dfrac{d \overline{u}}{d \overline{x}} \dfrac{d \overline{R}}{d \overline{x}}  \right) + \dfrac{3}{2} \overline{u}^{2} \left( \dfrac{d \overline{R}}{d \overline{x}} \right) ^{2} + \dfrac{4 \overline{u}}{Re \overline{R}} \dfrac{d \overline{R}}{d \overline{x}} + \dfrac{2}{We} \left( \dfrac{1}{\overline{R}} - \dfrac{1}{\overline{R}^{3 \gamma}} \right) = -\dfrac{C_{p}}{2} - \dfrac{\sigma}{2}  \left( 1 - \dfrac{1}{\overline{R}^{3 \gamma}}\right) 
\\
 - \dfrac{dP_{v}}{d \overline{T}} \dfrac{L_{ev} \rho_{v}}{\rho_{l}h_{b}} \overline{u} \dfrac{d \overline{R}}{d \overline{x}} \nonumber
\end{multline}
\begin{center}
$\alpha(x,t) = \dfrac{\dfrac{4}{3} \pi \overline{\eta} \overline{R}^{3}}{\left(1+ \dfrac{4}{3} \pi \overline{\eta} \overline{R}^{3} \right) }$
\end{center}
\begin{multline}
\overline{v} \dfrac{\partial \overline{v}}{\partial \overline{x}} = \left\lbrace - \dfrac{1}{\left(\rho_{v} + \dfrac{1}{2} \rho_{l} \right) \overline{v} u_{0}^{2} } \right\rbrace \left\lbrace \dfrac{\partial P}{\partial \overline{x}} - \dfrac{1}{2} \rho_{l} u_{0}^{2} \overline{u} + \dfrac{3}{8} \dfrac{C_{D} \rho_{l}}{\overline{R}} u_{0}^{2} \vert\left.\left( \overline{v} -\overline{u} \right)\right\vert  \right\rbrace \nonumber
\end{multline}

\indent The above-governing equations were integrated numerically using a higher-order adaptive Runge-Kutta scheme known as the Dormand-Prince scheme. Details of the adaptive numerical methods are discussed in detail in the next section.
\\ \\
\textit{\textbf{Step3: Numerical solution procedure:}}

\begin{itemize}
\item 	Dormand-Prince is called adaptive as it attempts to optimize the interval size and reduce the computational cost. The adaptive techniques algorithm should know when the interval (h) size must be adjusted.
\item 	Adaptive methods work by comparing the solution of an Ordinary Differential Equation by two methods, say Euler $\&$ Heun’s methods\cite{22}. From the solution difference, we find the relative error ofthe  Euler method with respect to Heun’s method (as Heun’s method is better in comparison with Euler)\cite{22}.
\item 	Then, this Error is compared with the allowable error in our computation, and based on that, a scaling factor is found for updating the interval size at each step.
\end{itemize}

\textbf{General procedure for finding Scaling parameter for ‘h’ with an illustrated example}
\paragraph{}
For a given initial value problem,
\begin{center}
$y^{(1)}(t) = f (t, y(t))$\\
$y(t_{0}) = y_{0}$
\end{center}

\indent \textbf{Step1}: Finding the solution using any two methods of numerical integration. Here, in this case, let the solution be done using Euler and Heun’s method
\begin{center}
$K_{1} = f(t_{k}, y_{k})$ \\
$K_{2} = f(t_{k} + h, y_{k} + hK_{1})$ \\ 
$y_{tmp} = y_{k} + hK_{1}$\\
$z_{tmp} = y_{k} + h \dfrac{K_{1}+K_{2}}{2} $
\end{center}

\indent Where $y_{tmp}$  and $z_{tmp}$  are the solutions obtained from Euler and Heun’s methods, respectively.
\\ \\
\indent \textbf{Step2}: The difference between the two solutions, $\vert y_{tmp} - z_{tmp} \vert$  is the Error. This error obtained is the error in the Euler method compared to Heun’s method, and the error is of order $O(h^{2})$. Therefore, error can be written as a function of $ h^{2} $ 
\begin{equation} \label{eq:scalingparameter1}
\vert y_{tmp} - z_{tmp} \vert = C h^{2}
\end{equation}

\indent \textbf{Step3}: Substituting scaling factor. Scaling 'h' by some factor 's' 

\begin{equation} \label{eq:scalingparameter2}
\vert y_{tmp} - z_{tmp} \vert = C h^{2} = C(s h^{2})
\end{equation}

\indent  \textbf{Step4}: Comparing the Error with allowable error $\&$ finding 's' 
\begin{center}
$C(s h^{2} ) < \varepsilon_{abs}$
\end{center}

\indent The contribution of the maximum error at the k$^{th}$ step should be proportional to the width of the interval relative to the whole interval
\begin{center}
$C(s h )^{2} < \varepsilon_{abs} \dfrac{sh}{t_{f} - t_{0}}$
\end{center}
\begin{equation} \label{eq:scalingparameter3}
C(s h )^{2} =\dfrac{1}{2} \varepsilon_{abs} \dfrac{sh}{t_{f} - t_{0}} = \dfrac{\varepsilon_{abs} s h}{2 \left( t_{f} - t_{0} \right) }
\end{equation}
\begin{equation}\label{eq:scalingparameter4}
\Rightarrow s(C h^{2} ) =  \dfrac{\varepsilon_{abs} h}{2 \left( t_{f} - t_{0} \right) }
\end{equation}

\indent Putting equation \ref{eq:scalingparameter1} in \ref{eq:scalingparameter3}
\begin{center}
 $s \vert y_{tmp} - z_{tmp} \vert =  \dfrac{\varepsilon_{abs} h}{2 \left( t_{f} - t_{0} \right) }$
\end{center}
\begin{equation}\label{eq:scalingparameter5}
\Rightarrow  s  =  \dfrac{\varepsilon_{abs} h}{2 \left( t_{f} - t_{0} \right)\vert y_{tmp} - z_{tmp} \vert }
\end{equation}

Updating 'h',\\
\begin{center}
$h^{\ast} = s \times h$
\end{center}
\indent where, $h^{\ast}$ is the updated value of h.\\ \\
\indent Depending on the value of 's', the value of 'h' is updated as follows 
\begin{itemize}
\item If s $\geq$ 2, $\qquad h^{\ast} = 2 \times h$
\item If 1 $\leq$ s $<$ 2, $\qquad h^{\ast} = h$ 
\item If s $<$ 1, $\qquad h^{\ast} = \dfrac{h}{2}$ (and try again)
\end{itemize}

\textbf{Algorithm for Dormand Prince Adaptive method}
\paragraph{}
In the Dormand Prince adaptive method, RK4 and RK5 methods are compared generally

\textbf{Step1}: Finding the slopes $k_{1}$ to $k_{7}$.
\\
$k_{1} = h f\left( t_{k}, y_{k}\right) $ \\ \\
$k_{2} = h f\left( t_{k}+\dfrac{1}{5}h, y_{k}+\dfrac{1}{5}k_{1}\right) $ \\ \\
$k_{3} = h f\left( t_{k}+\dfrac{3}{10}h, y_{k}+\dfrac{3}{40}k_{1}+\dfrac{9}{40}k_{2}\right) $ \\ \\
$k_{4} = h f\left( t_{k}+\dfrac{4}{5}h, y_{k}+\dfrac{44}{45}k_{1}-\dfrac{56}{15}k_{2}+\dfrac{32}{9}k_{3}\right) $ \\ \\
$k_{5} = h f\left( t_{k}+\dfrac{8}{9}h, y_{k}+\dfrac{19372}{6561}k_{1}-\dfrac{25360}{2187}k_{2}+\dfrac{64448}{6561}k_{3}-\dfrac{212}{729}k_{4} \right) $ 
\\ \\
$k_{6} = h f\left( t_{k}+h, y_{k}+\dfrac{9017}{3168}k_{1}-\dfrac{355}{33}k_{2}-\dfrac{46732}{5247}k_{3}+\dfrac{49}{176}k_{4}-\dfrac{5103}{18656}k_{5} \right) $
\\ \\
$k_{7} = h f\left( t_{k}+h, y_{k}+\dfrac{35}{384}k_{1}+\dfrac{500}{1113}k_{3}+\dfrac{125}{192}k_{4}-\dfrac{2187}{6784}k_{5}+\dfrac{11}{84}k_{6} \right) $
\\ \\
\indent \textbf{Step2}: Finding the solution using RK4 and RK5 method. 
\\ \\
The solution by Runge-Kutta 4th order method is given by:
\begin{center}
$y_{k+1} =  y_{k} + \dfrac{35}{384}k_{1} + \dfrac{500}{1113}k_{3} + \dfrac{125}{192}k_{4} - \dfrac{2187}{6784}k_{5} + \dfrac{11}{84}k_{6}  $
\end{center}

\indent The solution by Runge-Kutta 5th order method is given by:
\begin{center}
$y_{k+1} = y_{k} + \dfrac{5179}{57600}k_{1} + \dfrac{7571}{16695}k_{3} + \dfrac{393}{640}k_{4} - \dfrac{92097}{339200}k_{5} + \dfrac{187}{2100}k_{6} + \dfrac{1}{40}k_{7} $
\end{center}
\indent \textbf{Step3}: The Error in the RK4 method is calculated by finding the difference between the two solutions. 
The error for the Dormand Prince method is as shown below:
\begin{center}
$\left\vert z_{k+1} -  y_{k+1} \right\vert = \left\vert \dfrac{71}{57600}k_{1} -  \dfrac{71}{16695}k_{3} + \dfrac{71}{1920}k_{4} - \dfrac{17253}{339200}k_{5} + \dfrac{22}{525}k_{6} - \dfrac{1}{40}k_{7} \right\vert$
\end{center}

\indent \textbf{Step4}:  Finding the scaling factor and optimum interval size. The scaling factor for Dormand Prince is given by
\begin{center}
$s = \left( \dfrac{\varepsilon h}{ 2 \left\vert z_{k+1} -  y_{k+1} \right\vert  } \right)^{\dfrac{1}{5}}$
\end{center}
\indent The optimum interval size is given by Error prediction methods by Shampine\cite{22}
\begin{center}
$h_{opt} = sh$
\end{center}

\indent The above algorithm is incorporated in Matlab coding, and the model was solved for cases involving water and liquid nitrogen. 

\end{document}